\documentclass[aps,prx,reprint,superscriptaddress,amsmath,amssymb]{revtex4-2}

\usepackage[T1]{fontenc}
\usepackage[utf8]{inputenc}
\usepackage{color}
\usepackage{graphicx}
\usepackage{dcolumn}
\usepackage{bm}
\usepackage{hyperref}
\usepackage{amsthm}
\usepackage{braket}
\theoremstyle{plain}

\newtheorem*{claim}{Claim}

\begin{document}

\title{Convection Patterns in Nonequilibrium Kawasaki Dynamics at Low Temperature}

\author{Meander Van den Brande}
\affiliation{Department of Mathematics, King’s College London, Strand, London WC2R 2LS, United Kingdom}
\author{Kyosuke Adachi}
\affiliation{RIKEN Center for Interdisciplinary Theoretical and Mathematical Sciences, 2-1 Hirosawa, Wako 351-0198, Japan}
\affiliation{Nonequilibrium Physics of Living Matter Laboratory, RIKEN Pioneering Research Institute, 2-1 Hirosawa, Wako 351-0198, Japan}
\author{François Huveneers}
\affiliation{Department of Mathematics, King’s College London, Strand, London WC2R 2LS, United Kingdom}

\date{\today}

\begin{abstract}
We study a conservative stochastic lattice gas (Kawasaki dynamics) coupled in the bulk to a heat bath, which leads to standard phase separation at low uniform temperatures. 
Instead, a macroscopic temperature gradient drives the system into a nonequilibrium steady state.
In this state, the usual long-range order is replaced by robust convection patterns, featuring regularly spaced stripe structures. 
We show that these nonequilibrium states differ markedly from equilibrium configurations with the same local temperature profiles. 
Finally, we develop a macroscopic description that captures these behaviors and provides a unified framework for understanding the observed patterns.
\end{abstract}


\maketitle












\section{Introduction}\label{sec: introduction}

Nonequilibrium stationary states (NESS) are central to many-body physics, yet their structure is far less understood than that of equilibrium states, see  e.g.~\cite{Schmittmann1995,derrida1998,Bertini2002MFT,Seifert2012,Maes2020}. 
Because they do not arise from free-energy minimization, even simple driven systems may exhibit behaviors with no equilibrium counterpart. 
This is especially true when a conserved quantity is present, so that a steady state must satisfy a continuity equation. 
This constraint allows for long-range correlations, self-organized criticality~\cite{Grinstein_1990,Garrido_et_al_1990,Tasaki_2004,Adachi_Nakano_2024,Nakano_Adachi_2024}, and many of the collective phenomena observed in active matter~\cite{Vicsek_1995,Ramaswamy_2010,Marchetti_2013,Cates_Tailleur_2015}. 
Considerable progress has been made in systems that stay in local equilibrium, particularly in boundary-driven settings, with a very detailed description of the nonequilibrium fluctuations in some systems, see e.g.~\cite{EyinkLebowitzSpohn1990Hydrodynamics,Bodineau_derrida_2004,Bertini_et_al2015,mallick_2022,JonaLasinio2023,Bodineau2025}. 

In this work we investigate a conservative lattice gas in contact everywhere with a heat bath (Kawasaki dynamics), whose temperature varies smoothly across the system on a macroscopic scale. 
Particles interact through an Ising Hamiltonian, which at low and uniform temperature produces a symmetry-broken, phase-separated state. 
The imposed temperature gradient drives the system out of equilibrium by breaking detailed balance weakly, with a strength that scales inversely with the system size. 
This setting raises the fundamental question: how does a symmetry-broken phase respond to a weak nonequilibrium drive that preserves local equilibrium?

\begin{figure*}[t]
    \centering
    \includegraphics[width=0.9\linewidth]{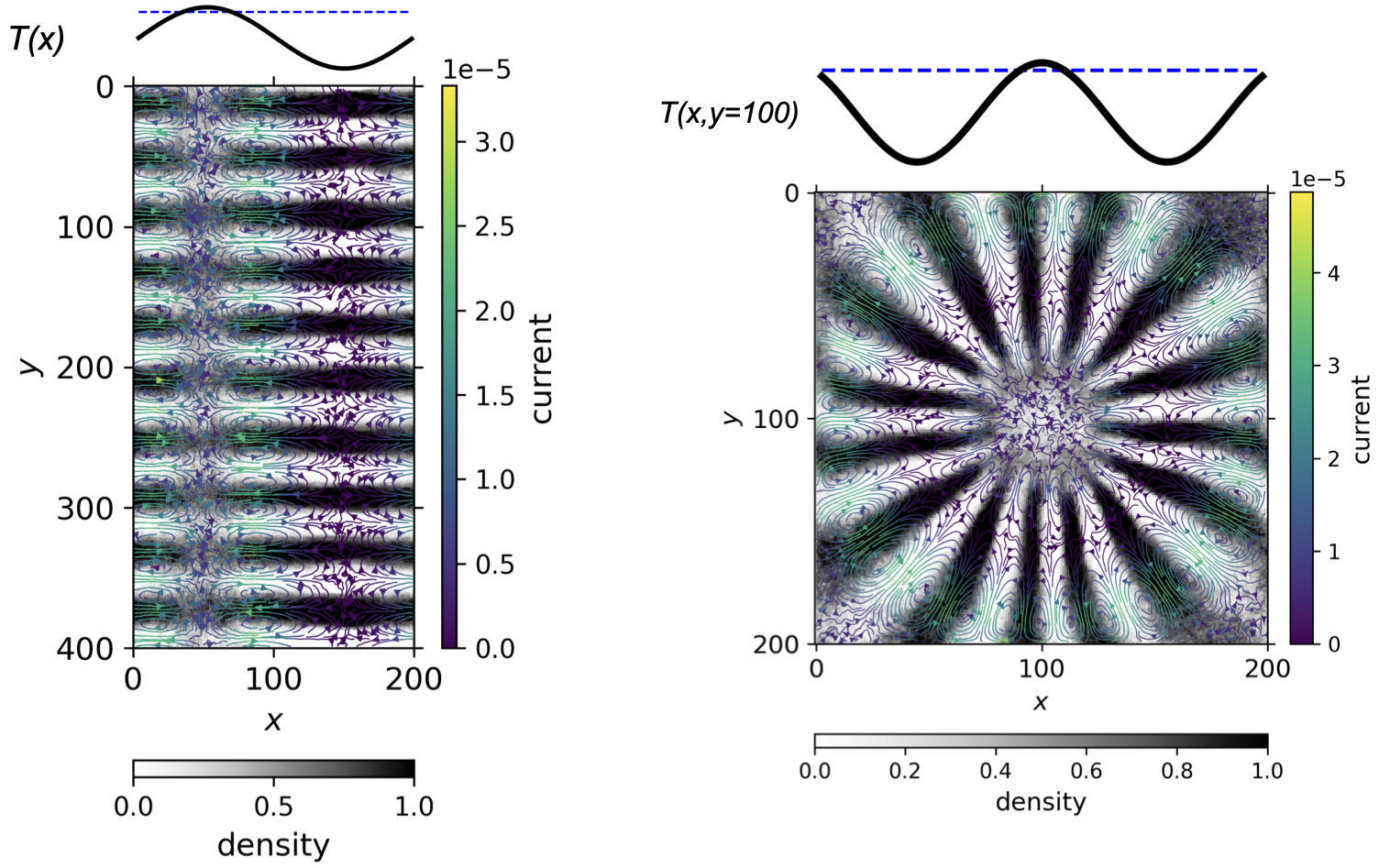}
    \caption{
    Time-averaged densities (grayscale) and particle currents (colored lines) for a single realization. 
    Averaging is performed over the final 25\% of the total simulation time ($t = 5 \times 10^{7}$). 
    Left: $x$-dependent temperature profile as in Eq.~\eqref{eq: x temperature profile} with $L_y = 2L_x = 400$; the profile is shown in the upper inset, where dotted lines indicate the critical temperature $T_c \approx 0.57$. 
    Right: Mexican-hat radial temperature profile as in Eqs.~\eqref{eq: radial temperature profile} and \eqref{eq: first radial profile} with $L_x = L_y = 200$;  $T(x,y=100)$ is shown in the upper inset, with the dotted line again representing $T_c$. 
    In both panels, $T_{\mathrm{mean}} = 0.4$, $T_{\mathrm{amp}} = 0.2$, and the filling factor is $\overline\rho = 1/2$.
    }
    \label{fig: densities and currents}
\end{figure*}

We find that the system develops convection-driven structures that fundamentally reorganize the macroscopic low-temperature phase. 
While the steady state is locally consistent with a Gibbs measure at the corresponding temperature, it globally forms a regular array of convection cells manifesting as density stripes. 
See Fig.~\ref{fig: densities and currents}. 
These cells appear in the subcritical region where equilibrium would produce complete phase separation, and their number increases with system size, thereby destroying conventional long-range order. 
The resulting phenomenology is closely related to Rayleigh–Bénard convection~\cite{chandrasekhar_1961,Cross_Hohenberg_1993,Getling_1998,Stevens_Hartmann_Verzicco_Lohse_2024}, Turing-type pattern formation~\cite{Turing_1952}, and dissipative structures more broadly~\cite{PrigogineLefever1968,Prigogine1969,NicolisPrigogine1977}. 
Its essential physical ingredients are local equilibrium, phase separation, and diffusive particle transport. 
At low temperature, this transport occurs within the bulk of the high and low-density phases (see the current lines in Fig.~\ref{fig: densities and currents}), where density fluctuations, or defects, propagate diffusively.


Previous studies of phase-separating systems driven out of equilibrium include models with a uniform macroscopic drive~\cite{katz_lebowitz_spohn_83,katz_lebowitz_spohn_84}, where domains elongate along the drive, or the driven Widom–Rowlinson gas~\cite{Dickman_2018,Zia_2025}, which exhibits stripes perpendicular to it. 
Related behavior is found in systems under shear flow~\cite{PhysRevLett.101.067203, PhysRevE.82.021126} or directed currents~\cite{PhysRevLett.109.130601}; in these cases, tangential fluxes along the interfaces suppress fluctuations.
Additionally, Kawasaki dynamics with antiferromagnetic interactions and a uniform drive was studied in Ref.~\cite{PhysRevLett.71.565}, where distinct nonequilibrium patterns were observed.

Other approaches have focused on the competition between different local rules, such as the simultaneous presence of Glauber and Kawasaki dynamics~\cite{PhysRevLett.59.1934, PhysRevA.40.6643, PhysRevE.110.024315}, or the linear instabilities appearing in non-reciprocal Ising models~\cite{10.21468/SciPostPhys.20.1.005}. 
Further, boundary-driven perturbations of Kawasaki dynamics have shown that coupling to magnetization reservoirs can induce uphill diffusion~\cite{Giardina_2018,Colangeli_2019,Giardina_2020,ColangeliGibertiVernia2023}.

Notably, our results are consistent with Refs.~\cite{Pleimling_2010,Li_2012}, which examined a set-up close to ours consisting of two subsystems separated by a sharp interface across which the temperature jumped from subcritical to infinite. 
In those studies, convection structures and density-modulated nonequilibrium stationary states were likewise observed. 
Similar stripe-like patterns were also found in thin films with a temperature gradient perpendicular to the film~\cite{Jaiswal2013}.

\subsection*{Summary of Main Results}

The model is described in detail in Sec.~\ref{sec: model}. 
The core of this work is an extensive numerical investigation of the steady state of the model, the results of which are presented in Sec.~\ref{sec: results}.
Our primary findings are as follows:

\paragraph{Robust convection cells.}
The system develops convection patterns reminiscent of Rayleigh-Bénard convection; these serve as the channel for heat transfer from the hot reservoir to the cold one. 
This is a robust feature of the steady state, appearing across various initial conditions and geometries (though not all); see Sec.~\ref{sec: convection patterns}. 
The cells form an approximately periodic arrangement that may become fuzzy due to fluctuations as the system size grows; see Sec.~\ref{sec: regularity patterns}. 
The number of cells increases with the linear system size $L$ for a fixed aspect ratio. 
Our data are consistent with the number of cells scaling as $L^{1/2}$, though fluctuations on time scales beyond our simulation limits might modify this figure; see Sec.~\ref{sec: scalcing number of stripes}.

\paragraph{Local equilibrium.}
The steady state of our nonequilibrium dynamics satisfies local equilibrium throughout the system: in the large-volume limit, the expectations of local observables are compatible with those of a Gibbs measure at the local temperature and some local chemical potential.
This property ultimately stems from the fact that detailed balance is only weakly broken, with violations in the transition rates being of order $1/L$; see Sec.~\ref{sec: local equilibrium}.

\paragraph{Diffusive transport.}
The convection cells in the steady state feature persistent particle currents. 
In Sec.~\ref{sec: scaling current}, we analyze how these currents scale with the system size $L$. 
Despite significant finite-size effects, the data are compatible with diffusive behavior, i.e., local currents scaling as $1/L$.

\begin{figure}[t]
     \centering
     \includegraphics[width=0.99\linewidth]{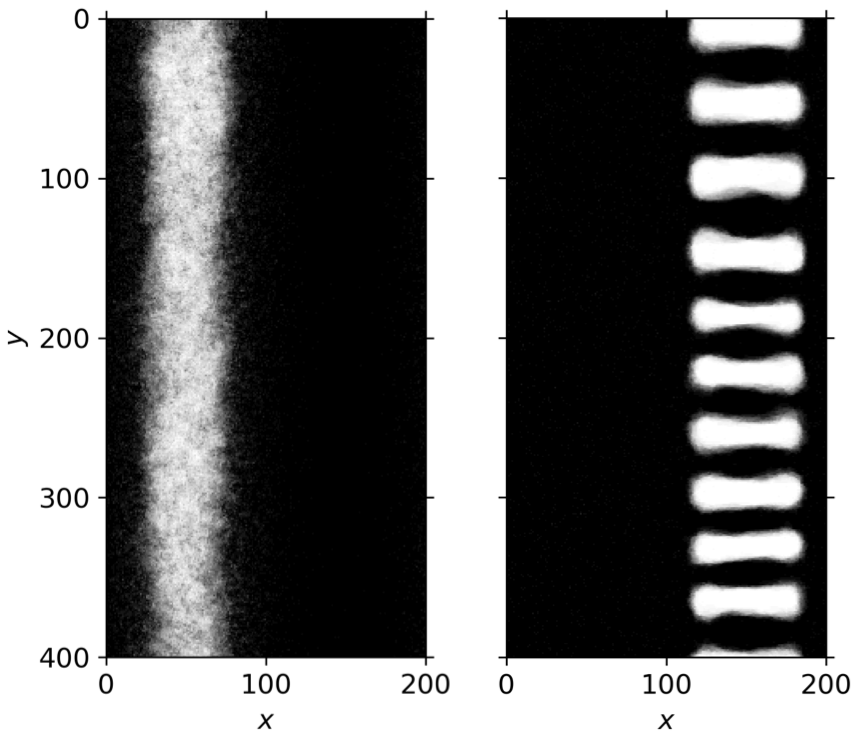}
     \caption{Equilibrium (left) and nonequilibrium (right) steady states.
    Time-averaged particle density at filling factor $\overline{\rho}=0.8$, obtained over the final $25\%$ of a simulation of duration $t = 5 \times 10^{7}$ for a single realization (grayscale as in Fig.~\ref{fig: densities and currents}). 
    The temperature varies along $x$ as in Eq.~\eqref{eq: x temperature profile}, with $T_{\mathrm{mean}}=0.4$ and $T_{\mathrm{amp}}=0.2$. 
    The system size is $L_y = 2L_x = 400$. 
    Equilibrium and nonequilibrium transition rates are given in Eqs.~\eqref{eq: equilibrium rates} and \eqref{eq: nonequilibrium rates}, respectively.}
     \label{fig:equil and non equil}
 \end{figure}

\paragraph{Chemical potential profiles (numerical).}
Most numerical results are obtained at half filling ($\overline{\rho} = 1/2$), where the data are compatible with a zero chemical potential across the system. 
The case $\overline{\rho} \ne 1/2$ is analyzed in Sec.~\ref{sec: other filling factors}, and the results are in sharp contrast with those expected from equilibrium. 
To highlight this, we compare the dynamics with an explicit equilibrium counterpart where the temperature is constant but the coupling strength varies spatially such that the ratio of coupling to temperature matches that of the nonequilibrium system.
The resulting steady states are markedly different; see Fig.~\ref{fig:equil and non equil}. 

In the second part of this paper, Sec.~\ref{sec: macroscopic description}, these findings are interpreted within the framework of a macroscopic theory that explicitly builds on the presence of local equilibrium and diffusive transport; see Sec.~\ref{sec: macroscopic constraints} for a precise statement of the underlying assumptions. 
The theory provides clarification on the following points:

\setcounter{paragraph}{0}

\paragraph{Chemical potential profiles (theoretical).}
It yields a clear and testable prediction for the chemical potential profile across the entire system. 
This prediction fits the data, specifically in Sec.~\ref{sec: other filling factors}, with discrepancies of only a few percent. 
These results are detailed in Sec.~\ref{sec: chemical potential profiles}.

\paragraph{Interface shape.}
It explains why the interface shape is no longer determined by mean-curvature considerations, but rather by nonequilibrium persistent currents that become dominant in the large-volume limit; see Sec.~\ref{sec: interface curvature}.

\paragraph{Fragmentation into convection cells.}
It shows that the number of convection cells must grow unbounded with the system size; see Sec.~\ref{sec: stripe formation}.

Finally, we note that an understanding based on the linearization of the mean-field dynamics has been developed in Ref.~\cite{mean_field_2026}. 
There, it is shown that the convective phase emerges as a dynamical instability, in a manner analogous to a Turing instability.

\section{Model}\label{sec: model}

We consider a stochastic lattice gas on a two-dimensional rectangular lattice $\Lambda$ with $L_x \times L_y$ sites and periodic boundary conditions. 
Each site $\mathbf x \in \Lambda$ is either empty, $n_{\mathbf x} = 0$, or occupied, $n_{\mathbf x} = 1$. 
A configuration is denoted by $\eta=(n_{\mathbf x})_{\mathbf x\in\Lambda}$. 
The dynamics conserves the total particle number $N$, and we denote the filling factor by $\bar\rho = N/V$ with $V=L_xL_y$. 
We study large systems by increasing $L_x$ up to 200 while keeping the aspect ratio $L_x/L_y$ fixed.

\subsection{Nonequilibrium Kawasaki Dynamics}\label{sec: Kawasaki out of equilibrium}

The energy of a configuration $\eta=(n_{\mathbf x})_{\mathbf x\in\Lambda}$ is given by the ferromagnetic Ising Hamiltonian
\begin{equation}\label{eq: ising energy}
E(\eta) = -J \sum_{\mathbf x \sim \mathbf y} (2n_{\mathbf x}-1)(2n_{\mathbf y}-1),
\end{equation}
with $J>0$ ($J=1/4$ in all simulations). 
Introducing spin variables $\sigma_{\mathbf x}=2n_{\mathbf x}-1=\pm1$ yields the standard Ising interaction.

The temperature varies on the macroscopic scale. 
Let $\mathsf T(u_x,u_y)$ be a smooth function on $[0,1]^2$, and define the microscopic temperature field as
\begin{equation*}
T(x,y)=\mathsf T(x/L_x,\,y/L_y).
\end{equation*}
Our main example is
\begin{equation}\label{eq: x temperature profile}
\mathsf T(u_x,u_y)=T_{\rm mean}+T_{\rm amp}\sin(2\pi u_x).
\end{equation}
We also consider radial profiles,
\begin{equation}\label{eq: radial temperature profile}
\mathsf T(u_x,u_y)=T_{\rm mean}+T_{\rm amp}\,\varphi(r)
\end{equation}
with $r^2=(u_x-0.5)^2+(u_y-0.5)^2$ and $\int_{[0,1]^2}\varphi(r)\,du_xdu_y=0$. 
Periodic boundaries ensure continuity of $T$ throughout.

The dynamics is a continuous-time Kawasaki exchange process. 
For a nearest-neighbor bond $(\mathbf x,\mathbf y)$ and configuration $\eta$, let $\eta^{\mathbf x,\mathbf y}$ be the configuration obtained by exchanging $n_{\mathbf x}$ and $n_{\mathbf y}$. 
Define the bond temperature
\begin{equation*}
T_{\mathbf x,\mathbf y}=
T\left(\frac{\mathbf x + \mathbf y}{2}\right),
\end{equation*}
as well as the energy difference 
\begin{equation}\label{eq: energy difference}
    \begin{aligned}
        &\Delta E_{\mathbf x,\mathbf y} = E(\eta^{\mathbf x,\mathbf y})-E(\eta)
    = -4J (n_{\mathbf y} - n_{\mathbf x}) \Delta_{x,y},\\
     &
     \text{with }\Delta_{x,y} = 
    \sum_{\mathbf z: \mathbf z \sim \mathbf x,\mathbf z \ne \mathbf y} n_{\mathbf z} 
    - 
    \sum_{\mathbf z: \mathbf z \sim \mathbf y,\mathbf z \ne \mathbf x} n_{\mathbf z}.
    \end{aligned}
\end{equation}
The transition rate is
\begin{equation}\label{eq: nonequilibrium rates}
W(\eta\to\eta^{\mathbf x,\mathbf y})
=\frac{\gamma}{1+e^{\Delta E_{\mathbf x,\mathbf y}/T_{\mathbf x,\mathbf y}}}.
\end{equation}
Spatial variations in $T$ break detailed balance.
To demonstrate the robustness of the phenomenology, we explore alternative dynamics in Appendix~\ref{sec: other dynamics}, including other temperature profiles and temperature-dependent rate $\gamma$.

\subsection{Continuity Equation}\label{sec: current computation}

Conservation of particle number and the restriction of nonzero transition rates to nearest neighbors imply that the stochastic evolution of \( n_{\mathbf x} \) can be written as
\begin{equation}\label{eq: lattice divergence form}
\frac{d n_{\mathbf x}}{dt}
= -\bigl( \mathsf j_{\mathbf x,\mathbf x+\mathbf e_1} - \mathsf j_{\mathbf x-\mathbf e_1,\mathbf x}
+ \mathsf j_{\mathbf x,\mathbf x+\mathbf e_2} - \mathsf j_{\mathbf x-\mathbf e_2,\mathbf x} \bigr)
\end{equation}
with  \( \mathbf e_1=(1,0) \) and \( \mathbf e_2=(0,1) \).
Here, $\mathsf j_{\mathbf x,\mathbf y}$ denotes the stochastic particle current, that can be decomposed as
\[
    \mathsf j_{\mathbf x, \mathbf y} = 
    j_{\mathbf x, \mathbf y} + \xi_{\mathbf x, \mathbf y}
\]
where $j_{\mathbf x, \mathbf y}$ is an observable, i.e.\@ a function of the configuration, and $\xi_{\mathbf x,\mathbf y}$ is a stochastic noise. 


Let us derive an expression for the current $j_{\mathbf x,\mathbf y}$.
The stochastic evolution of a generic observable $A(\eta)$ is given by
\[
    \frac{dA(\eta)}{dt} 
    = 
    \mathcal L A(\eta) + \xi_{A,\eta},
\]
where $\xi_{A,\eta}$ is a noise (a martingale increment) and where $\mathcal L$ is the generator of the dynamics, given by
\[
    \mathcal L A(\eta) 
    = 
    \sum_{\eta'} \big(A(\eta') - A(\eta)\big) W(\eta \to \eta').
\]
This is Dykin's martingale formula, see e.g.\@ Ref.~\cite{kallenberg2021}.
Taking $A = n_{\mathbf x}$ yields
\[
\mathcal L n_{\mathbf x}(\eta) 
= 
\sum_{\mathbf y: \mathbf y \sim \mathbf x}
\big(n_{\mathbf x}(\eta^{\mathbf x,\mathbf y}) - n_{\mathbf x}(\eta)\big)
W(\eta \to \eta^{\mathbf x,\mathbf y}).
\]
Using
\[
    \frac{1}{1 + e^x} = \frac{1}{2}\left( 1 - \tanh\frac{x}{2} \right)
\]
and the expression~\eqref{eq: nonequilibrium rates} for the rates, we obtain
\begin{equation}\label{eq: L nx 1st expression}
\mathcal L n_{\mathbf x} 
= 
\frac{\gamma}{2}
\sum_{\mathbf y: \mathbf y \sim \mathbf x}
(n_{\mathbf y} - n_{\mathbf x})
\left( 1 - \tanh\frac{\Delta E_{\mathbf x,\mathbf y}}{2 T_{\mathbf x,\mathbf y}} \right),
\end{equation}
where we omit the explicit dependence on $\eta$, knowing that $\eta = (n_{\mathbf x})_{\mathbf x}$.
Because $\Delta E_{\mathbf x,\mathbf y}$ depends on wether a particle moves from $\mathbf x$ to $\mathbf y$ or vice versa, we decompose
\begin{align*}
    &(n_{\mathbf y} - n_{\mathbf x})
    \tanh\frac{\Delta E_{\mathbf x,\mathbf y}}{2 T_{\mathbf x,\mathbf y}}\\
     &= 
     n_{\mathbf y}(1 - n_{\mathbf x})
     \tanh\frac{\Delta E_{\mathbf x,\mathbf y}}{2 T_{\mathbf x,\mathbf y}}
     - n_{\mathbf x}(1 - n_{\mathbf y})
     \tanh\frac{\Delta E_{\mathbf x,\mathbf y}}{2 T_{\mathbf x,\mathbf y}}\\
     &= 
    \big(n_{\mathbf y}(1 - n_{\mathbf x}) 
    + n_{\mathbf x}(1 - n_{\mathbf y})\big)
    \tanh\frac{2 J \Delta_{\mathbf x,\mathbf y}}{ T_{\mathbf x,\mathbf y}}
\end{align*}
using Eq.~\eqref{eq: energy difference}.
Inserting this in Eq.~\eqref{eq: L nx 1st expression} yields
\[
    \mathcal L n_{\mathbf x}
    = - (j_{\mathbf x,\mathbf x+\mathbf e_1} - j_{\mathbf x-\mathbf e_1,\mathbf x})
     - (j_{\mathbf x,\mathbf x+\mathbf e_2} - j_{\mathbf x-\mathbf e_2,\mathbf x})
\]
with 
\begin{multline}\label{eq: nonequilibrium current}
j_{\mathbf x,\mathbf y}
= 
- \frac{\gamma}{2} (n_\mathbf y - n_\mathbf x) \\
+ 
\frac{\gamma}{2}
\big(n_{\mathbf y}(1 - n_{\mathbf x}) + n_{\mathbf x}(1 - n_{\mathbf y})\big)
     \tanh\frac{2J\Delta_{\mathbf x,\mathbf y}}{T_{\mathbf x,\mathbf y}}.
\end{multline}



\subsection{Inhomogeneous Equilibrium Dynamics}\label{sec: equilibrium dynamics}

For comparison, we introduce a related process that satisfies detailed balance; its rates differ from the nonequilibrium process described above only by corrections of order $1/L$.
Define the modified energy
\begin{equation}
E_T(\eta)
=-J\sum_{\mathbf x\sim\mathbf y}\frac{1}{T_{\mathbf x,\mathbf y}}
(2n_{\mathbf x}-1)(2n_{\mathbf y}-1),
\end{equation} 
where $-J/T_{\mathbf x, \mathbf y}$ now plays the role of a spatially varying coupling strength. 
The Kawasaki rates are
\begin{equation}\label{eq: equilibrium rates}
W_{\rm eq}(\eta\to\eta^{\mathbf x,\mathbf y})
=\frac{\gamma}{1+e^{\Delta E_{T,\mathbf x,\mathbf y}}}
\end{equation}
with $\Delta E_{T,\mathbf x,\mathbf y}$ given by 
\begin{equation}\label{eq: energy difference equilibrium}
    \begin{aligned}
        &\Delta E_{\mathbf x,\mathbf y} = E_T(\eta^{\mathbf x,\mathbf y})-E_T(\eta)
    = -4J (n_{\mathbf y} - n_{\mathbf x}) \Delta_{x,y}^T,\\
     &
     \text{with }\Delta_{x,y}^T = 
    \sum_{\mathbf z: \mathbf z \sim \mathbf x,\mathbf z \ne \mathbf y} 
    \frac{n_{\mathbf z}}{T_{\mathbf x,\mathbf z}} 
    - 
    \sum_{\mathbf z: \mathbf z \sim \mathbf y,\mathbf z \ne \mathbf x} 
    \frac{n_{\mathbf z}}{T_{\mathbf y, \mathbf z}}.
    \end{aligned}
\end{equation}
This process is conservative and the continuity equation~\eqref{eq: lattice divergence form} still holds, provided the current $j_{\mathbf x,\mathbf y}$ is now replaced by 
\begin{multline}\label{eq: equilibrium current}
j_{\mathbf x,\mathbf y}^T
= 
- \frac{\gamma}{2} (n_\mathbf y - n_\mathbf x) \\
+ 
\frac{\gamma}{2}
\big(n_{\mathbf y}(1 - n_{\mathbf x}) + n_{\mathbf x}(1 - n_{\mathbf y})\big)
     \tanh\left(2J\Delta^T_{\mathbf x,\mathbf y}\right).
\end{multline}

Crucially, because the rates in Eq.~\eqref{eq: equilibrium rates} define an equilibrium dynamics and because the process is conservative, the invariant states are Gibbs states of the type $e^{-(E_T - \mu N)}$, for some chemical potential $\mu$. 
Since our focus is on the stationary states, the dynamics is here merely a way to access this Gibbs state, rather than the fundamental object of study.




\section{Numerical Results}\label{sec: results}

We present the main conclusions of the numerical analysis of the nonequilibrium dynamics with rates given in Eq.~\eqref{eq: nonequilibrium rates}.
We work at filling factor $\overline\rho = 1/2$, except in Sec.~\ref{sec: other filling factors}, where we also use the equilibrium dynamics with rates given in Eq.~\eqref{eq: equilibrium rates} as a comparison point.
The initial state is either an infinite-temperature (white noise) configuration, obtained by placing $N=\bar\rho L_x L_y$ particles uniformly at random on the lattice, or a phase-separated configuration of prescribed geometry.
We have examined cases in which the temperature profile crosses the critical point as well as cases that remain entirely subcritical. 
Although finite-size scaling is more difficult to analyze when the critical point is crossed, the qualitative phenomenology appears to be the same in both cases.

\subsection{Simulation Method and Convergence to Stationarity}\label{sec: simulation and convergence}
We simulate the dynamics using a standard kinetic Monte-Carlo scheme. 
One Monte-Carlo cycle consists of the following steps:
\begin{enumerate}

\item Choose a particle at random.

\item Among the unoccupied nearest neighboring sites, select one at random with probability $W\Delta t$, where $W$ is the transition rate given in Eqs.~(\ref{eq: nonequilibrium rates}) and (\ref{eq: equilibrium rates}), and where $\Delta t = 1/(4\gamma)$. 
If a site is selected, move the particle; otherwise it remains in place.
The choice $\Delta t = 1/(4\gamma)$ guarantees that the probability is independent of the value of $\gamma$ and that the probability of moving at all never exceeds unity.

\item Repeat steps 1–2 a total of $N$ times and then increment time by $1$ unit.
\end{enumerate}
Time is measured in units of $1/\gamma$, with $\gamma$ as in Eq.~(\ref{eq: nonequilibrium rates}) and (\ref{eq: equilibrium rates}), so the parameter $\gamma$ plays no further role.

Typical simulations run up to $t = 5 \times 10^7$, a duration sufficient to reach a state that exhibits the main characteristics of true stationarity. 
The strongest evidence for this convergence is that significantly different initial conditions lead to similar states on these time scales. 
For instance, the dynamical process shown in the top panel of Fig.~\ref{fig: snapshots evolution}, initiated from a vertically segregated state, yields a final configuration comparable to those starting from horizontally segregated or white-noise initial conditions. 
Additional standard stationarity probes are detailed in Appendix~\ref{sec: evidence for stationarity}. 
We note, however, that once convection cells form, they become remarkably robust. 
On much longer time scales, the number of cells is expected to fluctuate, a process too slow to be witnessed in our simulations.
Consequently, the scaling of the number of convection cells with system size (Sec.~\ref{sec: scalcing number of stripes}) is a result that could be affected by these possible long-term dynamics.


\subsection{Convection Patterns}\label{sec: convection patterns}
\begin{figure*}[t]
     \centering
     \includegraphics[width=0.99\textwidth]{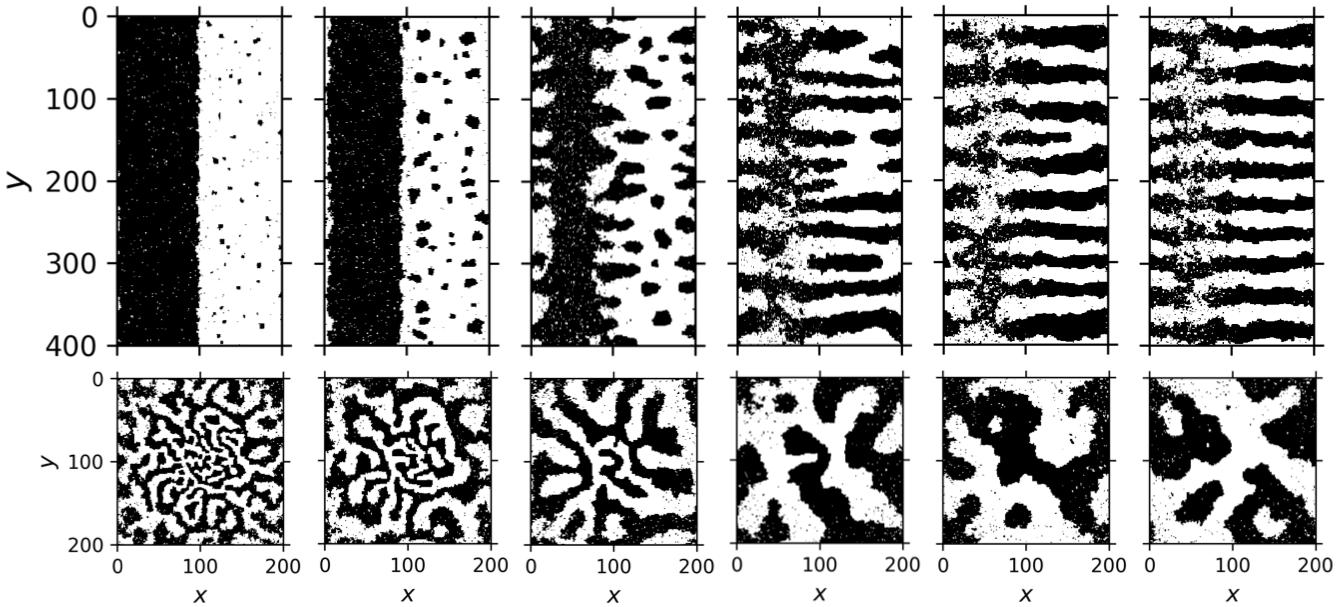}


     
     \caption{Time evolution of the density field.
Snapshots of the particle density at $0.1\%,\,0.4\%,\,1.6\%,\,6.3\%,\,25\%,$ and $100\%$ of the total simulation time $t=5\times10^{7}$, for a single realization at filling factor $\overline{\rho}=1/2$. 
Upper row: $x$-dependent temperature profile as in Eq.~\eqref{eq: x temperature profile} with $T_{\mathrm{mean}}=0.4$, $T_{\mathrm{amp}}=0.2$, and system size $L_y=2L_x=400$. 
Lower row: dip temperature profile as in Eqs.~\eqref{eq: radial temperature profile} and \eqref{eq: second radial profile} with $T_{\mathrm{mean}}=0.4$, $T_{\mathrm{amp}}=0.2$, and system size $L_x=L_y=200$. The system is initialized from an infinite-temperature (white-noise) configuration.}
     \label{fig: snapshots evolution}
 \end{figure*}

Consider the temperature profile of Eq.~\eqref{eq: x temperature profile}, with system sizes up to $L_x = 200$ and $L_y = 2L_x$. 
In the subcritical region, the dynamics generates regularly arranged convection cells manifesting as stripes of alternating high and low particle density. 
This occurs for all tested initial conditions: 
an infinite-temperature configuration, 
a phase-separated state with a single vertical interface at $x = L_x/2$, 
and a phase-separated state with one or several horizontal interfaces. 
See the left panel in Fig.~\ref{fig: densities and currents} and the upper panels of Fig.~\ref{fig: snapshots evolution}.

The convection cells are robust. 
Once formed, we never observe changes in the number of stripes for large systems. 
Such events arise only for very small lattices and become highly unlikely as the system size grows.
The position of the cells do fluctuate on somewhat shorter time scales, though.
Indeed, performing a time-average over longer times than in Fig.~\ref{fig: densities and currents} did reveal some more blurring for some realizations. 
This is consistent with the existence of a unique steady state that is invariant under translations along the $y$ direction.

This robustness is due to the persistent particle currents circulating within each cell, as illustrated in Fig.~\ref{fig: densities and currents}. 
The vector field shown there corresponds to the bond currents $(j_{\mathbf{x},\mathbf{x}+\mathbf{e}_1},\, j_{\mathbf{x},\mathbf{x}+\mathbf{e}_2})$, see Eq.~\eqref{eq: nonequilibrium current}, time-averaged over a certain period.
In practice, it is computed by counting the number of particle jumps from $\mathbf{x}$ to $\mathbf{y}$ minus those from $\mathbf{y}$ to $\mathbf{x}$ over a fixed integration time, and dividing by that time.
Averages are taken over the same time window used for the density profiles.
This window is chosen large enough to remove fluctuations but short enough so that the circulating currents do not reverse as the pattern move due to the eventual restoration of translational symmetry.
The presence of these currents is consistent with basic thermodynamics: 
energy is absorbed from the bath in the hot region, through the creation of domain walls, and released into it in the cold one through their destruction. 
The constant flux of particles makes this process possible. 



The emergence of periodic conduction patterns depends partly on the temperature profile in Eq.~\eqref{eq: x temperature profile}.
To assess this dependence, we consider alternative radial profiles such as Eq.~\eqref{eq: radial temperature profile} (see also Appendix~\ref{sec: other dynamics}).
We first examine
\begin{equation}\label{eq: first radial profile}
\varphi_1(r) =
\begin{cases}
\cos(2\pi \alpha_1 r), & \alpha_1 r \le 1,\\
1, & \alpha_1 r \ge 1,
\end{cases}
\end{equation}
with $\alpha_1 \simeq 0.88$, which ensures that $\varphi_1$ has zero average.
Simulations are performed with $L_x = L_y = 200$.
This profile produces a Mexican-hat–like temperature landscape: the temperature peaks at the center, decreases at intermediate $r$, and rises again at larger $r$.
The resulting time-averaged density and current fields, displayed on the lower panel of Fig.~\ref{fig: densities and currents}, appear as a distorted version of the reference pattern on the upper panel, indicating that the same phenomenology is at play.

A qualitatively different behavior arises for
\begin{equation}\label{eq: second radial profile}
\varphi_2(r) =
\begin{cases}
-\cos(2\pi \alpha_2 r), & \alpha_2 r \le 1/2,\\
1, & \alpha_2 r \ge 1/2,
\end{cases}
\end{equation}
with $\alpha_2 \simeq 0.32$.
This profile features a single central dip.
Here no stable pattern is formed.
Snapshots are shown on the lower panel of Fig.~\ref{fig: snapshots evolution}. 
They feature transient, star-like distortions along radial directions, and time averaging yields much faster homogenization than in the other cases.


\subsection{Regularity of the Convection Patterns}\label{sec: regularity patterns}

The convection patterns form a regular, approximately periodic arrangement; see Figs.~\ref{fig: densities and currents}, \ref{fig:equil and non equil}, and \ref{fig: snapshots evolution}.
A natural explanation is that each stripe carries a particle current, and current conservation requires that these currents balance at interfaces.
Stripes of equal width satisfy this condition automatically.
However, as the system size increases, fluctuations may cause distant stripes to acquire different widths, potentially degrading global periodicity.

To quantify this, following the approach of Ref.~\cite{Li_2012}, we compute the structure factor
\[
    S(\mathbf k) = \frac{1}{L_x L_y} \langle |\widetilde n(\mathbf k)|^2 \rangle, 
    \quad 
    \widetilde n(\mathbf k) = \sum_{\mathbf x} e^{2i\pi \mathbf k \cdot \mathbf x}(n(\mathbf x) - \overline\rho),
\]
with filling factor $\overline{\rho}$, wavevectors $\mathbf k=(k_x,k_y)$ given by $k_z= m_z/L_z$ ($m_z=0,\dots,L_z-1$ for $z=x,y$), and $\langle\cdot\rangle$ a form of steady-state average chosen to preserve the convection structure while suppressing short-scale fluctuations.
For a homogeneous configuration, $S(\mathbf k)=\mathcal O(1)$; for a periodically ordered state, $S(\mathbf k_{\mathrm{peak}})\sim V$ with $V=L_xL_y$, indicating long-range periodic order at $\mathbf k_{\mathrm{peak}}$.

\begin{figure}[t]
    \centering
    \includegraphics[width=0.95\linewidth]{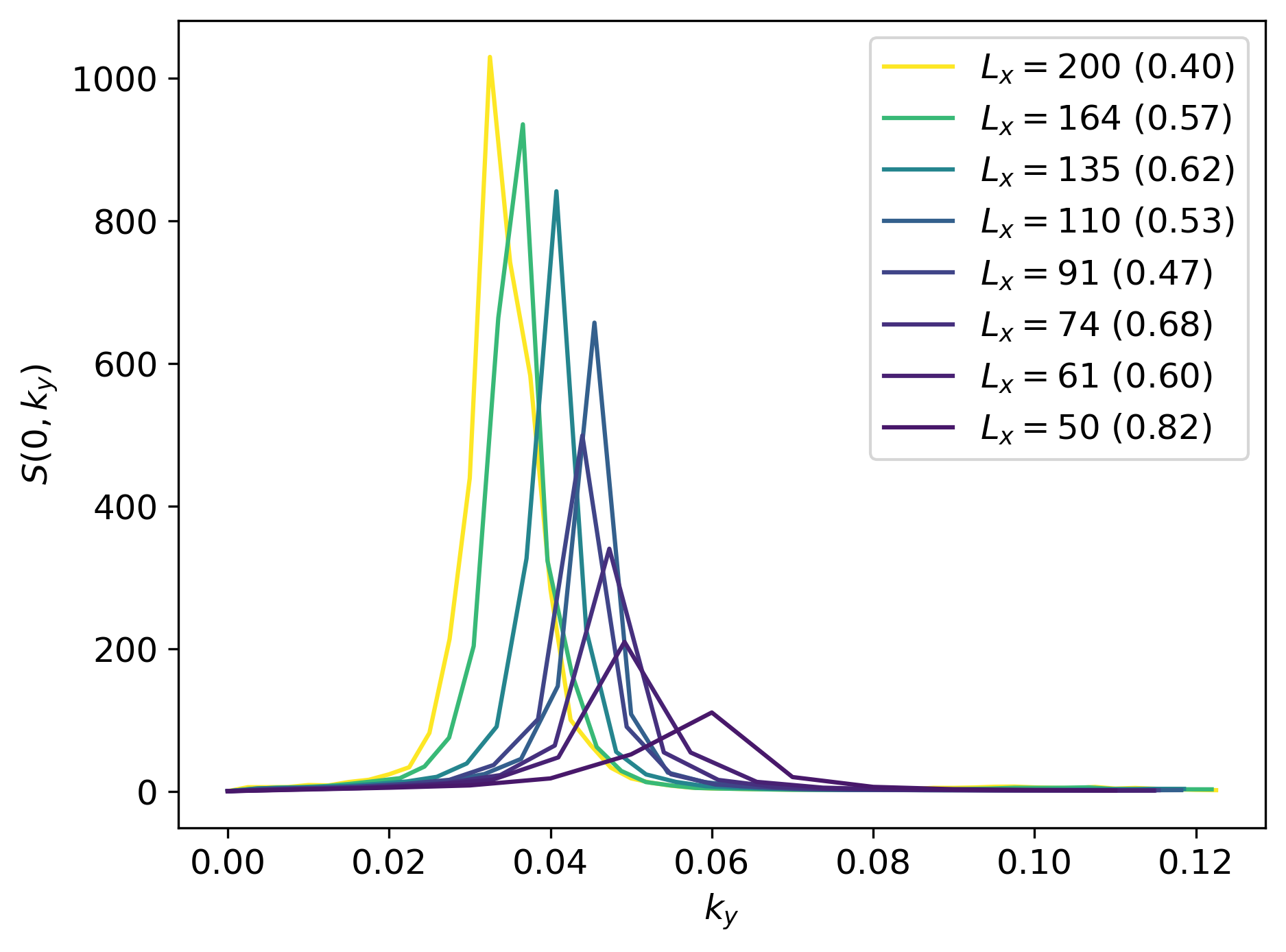}
    \caption{Structure factor.
    Structure factor $S(0,k_y)$ as a function of $k_y$ for several system sizes $L_x$, at fixed aspect ratio $L_y/L_x=2$ and filling factor $\overline{\rho}=1/2$. 
    The temperature varies along $x$ as in Eq.~\eqref{eq: x temperature profile}. 
    The value of $|\tilde n(\mathbf k)|^2$ is averaged over the final $25\%$ of a simulation of duration $t=5\times10^{7}$, and over a proportion of $40$ realizations, indicated in parentheses for each $L_x$. }
    \label{fig: structure factor}
\end{figure}

Figure~\ref{fig: structure factor} shows $S$ as a function of $\mathbf k=(0,k_y)$ for $\overline{\rho}=1/2$ and the temperature profile of Eq.~\eqref{eq: x temperature profile} with $T_{\mathrm{mean}}=0.4$ and $T_{\mathrm{amp}}=0.2$, for various system sizes and given aspect ratio $L_y/L_x = 2$.
To avoid washing out the signal, we do not average over realizations with different numbers of cells at a given value of $L_x$.
Instead, for each realization we sample $25$ configurations equally spaced in time over the last $25\%$ of the simulation and determine, for that realization, the most frequently occurring number of cells (typically overwhelmingly dominant, with rare deviations due to short-lived bridges between neighboring cells). 
This assigns a unique cell number to each realization. 
We then consider 40 realizations and retain only those whose assigned cell number matches the most probable value across realizations. 
Finally, \(S(\mathbf k)\) is averaged over all $25$ sampled configurations of the retained realizations, and \(S(0,k_y)\) is reported.

Two features emerge.
First, $S(\mathbf k)$ exhibits a clear peak at $k_{y,\mathrm{peak}}$, with $k_{y,\mathrm{peak}}\to 0$ as $L_x$ increases, consistent with the sublinear growth of the cell number.
Second, although $S(\mathbf k_{\mathrm{peak}})$ increases with system size, it grows slower than $V$.
This indicates that periodicity may be visible over intermediate distances, but true periodic long-range order is eventually destroyed by fluctuations.

\subsection{Scaling of the Number of Convection Cells}\label{sec: scalcing number of stripes}

We investigate how the number of convection cells, or equivalently of high-density stripes $\mathcal N$, scales with system size at fixed aspect ratio.
We restrict ourselves to the stripes that appear in the low temperature region for the temperature profile in Eq.~\eqref{eq: x temperature profile} with filling factor $\overline\rho = 1/2$.
Some remarks are useful to frame the problem.
First, one expects a scaling $\mathcal N \sim L^a$ with $0<a<1$, as both $a=0$ and $a=1$ are incompatible with local equilibrium, as will be discussed in Sec.~\ref{sec: local equilibrium}.
Other forms, such as $\mathcal N \sim \log L$, cannot be excluded a priori but are not supported by our data.
Second, the dynamics admits a relatively broad range of stripe numbers that remain stable on simulation timescales.
Starting from perfectly striped configurations with $1\le \mathcal N \le 24$, we find that they give rise to stable patterns with the same number of stripes for at least $8\le \mathcal N \le 17$ during the simulation time $t=5\times 10^7$. 
Finally, as stressed already in Sec.\ref{sec: simulation and convergence}, it is possible that the figure would be modified if we could track the dynamics for much longer time scales. 

\begin{figure}[t]
    \centering
    \includegraphics[width=0.95\linewidth]{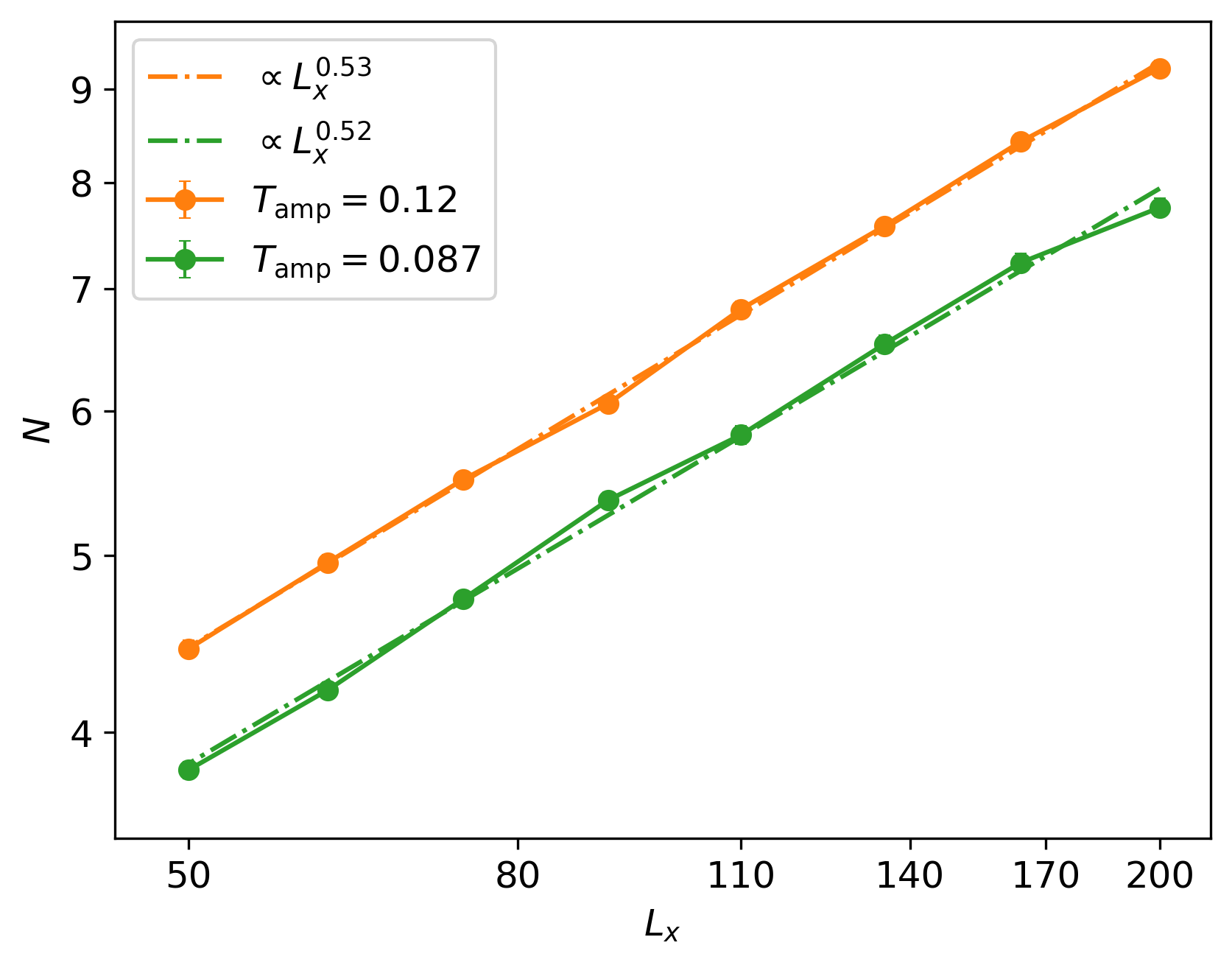}
    \caption{Number of high-density stripes $\mathcal N$ as a function of the system size $L_x$ for fixed aspect ratio $L_y/L_x$ and filling factor $\overline\rho = 1/2$. 
    The temperature varies along $x$ as in Eq.~\eqref{eq: x temperature profile} for $T_{\mathrm{mean}}=0.42$, and $T_{\mathrm{amp}}=0.12$ (orange) or $T_{\mathrm{amp}}=0.087$ (green). Average over 40 samples and over the last $25\%$ of the total time $t=1.33 \times 10^7$. See main text.}
    \label{fig: number of stripes}
\end{figure}

To extract the most intrinsic scaling, we initialize the system at infinite temperature (white noise).
Fig.~\ref{fig: number of stripes} shows results for $T_{\mathrm{mean}}=0.42$ with $T_{\mathrm{amp}}=0.12$ and $0.087$, ensuring that the temperature remains subcritical throughout the system.
The method used for counting stripes is reported in Appendix~\ref{sec: measuring number of stripes}.
The data are consistent with the scaling $\mathcal N \sim L_x^{1/2}$.

The interpretation of the data is complicated by a threshold effect arising from the fact that the number of stripes is an integer, which is not suppressed by increasing the number of samples.
To mitigate this, we simulate systems with aspect ratio $L_y/L_x = 6$, for $50 \le L_x \le 200$, instead of the ratio $L_y/L_x = 2$ used so far.
On the other hand, holding $L_x$ fixed, we verify that $\mathcal N$ grows linearly with $L_y$. 
The values reported in Fig.~\ref{fig: number of stripes} are obtained by dividing the measured stripe counts by~3, and are expected to provide a faithful representation of the average number of stripes for the aspect ratio $L_y/L_x=2$.

We have also examined cases where the temperature crosses the critical point.
Here exponents significantly larger than $1/2$ appear, depending on $T_{\mathrm{amp}}$, though the curves exhibit noticeable downward bending for large system sizes.
No universal exponent was found in this regime; the corresponding data are presented in Appendix~\ref{sec: scaling number stripes super ciritical}.

\subsection{Local Equilibrium}\label{sec: local equilibrium}

Local equilibrium means that the expectation of local observables coincides with their value in equilibrium at the local temperature and some local chemical potential in the large volume limit. 
The snapshots in the top row of Fig.~\ref{fig: snapshots evolution} furnish a first indication of local equilibrium: as defects migrate towards the cold part, they nucleate bigger clusters, consistent with the fact that, in equilibrium, the density of defects decreases with decreasing temperature.

\begin{figure}[t]
    \centering
    \includegraphics[width=0.99\linewidth]{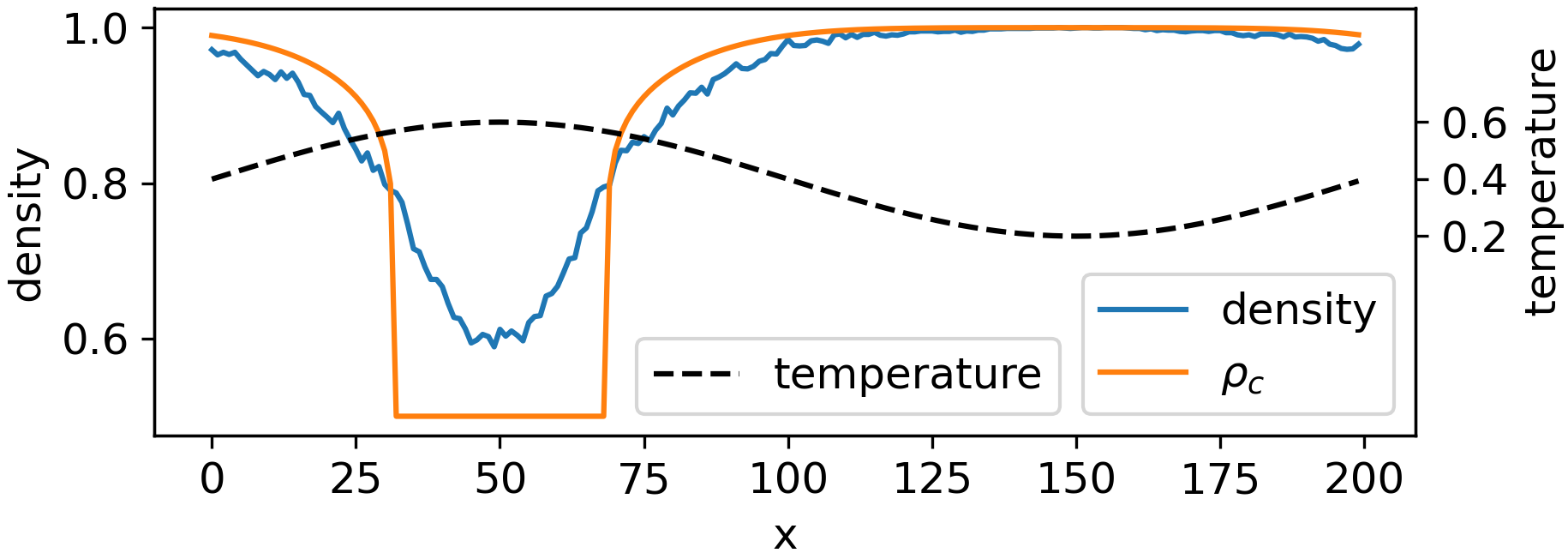}
    \caption{Time and $y$-averaged density within high density lines as a function of $x$.
    Temperature as in Eq.~\eqref{eq: x temperature profile} with $T_{\mathrm{mean}}=0.4$, $T_{\mathrm{amp}}=0.2$, 
    system size $L_y=2L_x=400$, $\overline\rho=1/2$. 
    Blue line: 
    time average taken over the final $75\%$ of a simulation of duration $t=5\times10^{7}$ and spatial average in the $y$ direction taken over a fraction of the sites with $y$-coordinate corresponding to maximal density at $x=L_x/2$ (see main text).
    Single realization, starting from an initial configuration with $13$ equally spaced stripes.
    Orange line: equilibrium spontaneous density at the local temperature.}
    \label{fig: local equilibrium density}
\end{figure}

Verifying local equilibrium in the cold region is delicate due to the proximity of interfaces. 
In Fig.~\ref{fig: local equilibrium density}, we consider the temperature profile of Eq.~\eqref{eq: x temperature profile} at filling factor \(\overline{\rho}=1/2\).
We compute the time-averaged density \(\langle n_{\mathbf x}\rangle\) over the interval specified in the figure caption. On the vertical line $x=L_x/2$, where the temperature is subcritical, we select \(1/8\) of the sites with the largest values of \(|2\langle n_{\mathbf x}\rangle-1|\), corresponding to densities close to 0 or 1, and retain only those with \(\langle n_{\mathbf x}\rangle \simeq 1\).
This procedure yields approximately \(1/16\) of the sites on the line. 
We then average the density over the \(y\) direction along the corresponding rows. These rows lie well within the bulk of the high-density stripes in the low-temperature region.

Since the chemical potential is expected to remain zero throughout the system (see Sec.~\ref{sec: chemical potential profiles}), we compare the measured densities with Onsager’s exact result~\cite{yang1952ising} for the equilibrium spontaneous density at the local temperature, $\rho_c(T(\mathbf x))\ge 1/2$. 
Overall agreement is good. 
Importantly, the density does not fall significantly below the spontaneous density, which would signal the presence of a metastable state.
Slight deviations to this are observed, but we checked that this effect decreases when sampling deeper inside the stripes (here restricting the average to the central $1/8$ of the high-density region), consistent with a finite-size effect. 
In the high-temperature region the density exceeds $1/2$. 
This does not violate global symmetry, since the average is restricted to high-density lines.
It is also compatible with local equilibrium, although it differs from the value expected at $\mu=0$. 
We attribute this discrepancy to finite-size effects and expect the condition $\mu=0$ to hold throughout in the thermodynamic limit, a point that warrants further investigation.

\subsection{Scaling of the Particle Current}\label{sec: scaling current}

At low temperature and within each phase, particle fluxes are driven by density variations, which in turn stem from temperature variations; see Fig.~\ref{fig: local equilibrium density}.
When $L_x$ increases at fixed macroscopic temperature profile, the microscopic temperature gradient scales as $1/L_x$ and, away from criticality, the local density variation also scales as $1/L_x$.
Given the diffusive transport in this system (Fick’s law, see Sec.~\ref{sec: macroscopic constraints}), the current is expected to be proportional to the density gradient, and to scale as $1/L_x$.

We use the temperature profile of Eq.~\eqref{eq: x temperature profile} at filling factor \(\overline{\rho}=1/2\) and measure the time-averaged current in the \(x\) direction across bonds intersecting \(x=L_x/2\), where the current is approximately maximal. 
As in the analysis of local equilibrium, care must be taken to restrict measurements to regions well inside the high and low-density stripes, where the current is negative and positive, respectively. 
Accordingly, we compute the time-averaged density \(\langle n_{\mathbf x}\rangle\) over a given time window and select the \(1/8\) of sites with the largest values of \(|2\langle n_{\mathbf x}\rangle - 1|\). 
The current through each horizontal bond originating from these sites is then averaged over the same time interval. 
Finally, we average these currents, assigning negative sign to those in high-density regions (\(\langle n_{\mathbf x}\rangle \simeq 1\)) and positive sign to those in low-density regions (\(\langle n_{\mathbf x}\rangle \simeq 0\)).

\begin{figure}[t]
    \centering
    \includegraphics[width=0.99\linewidth]{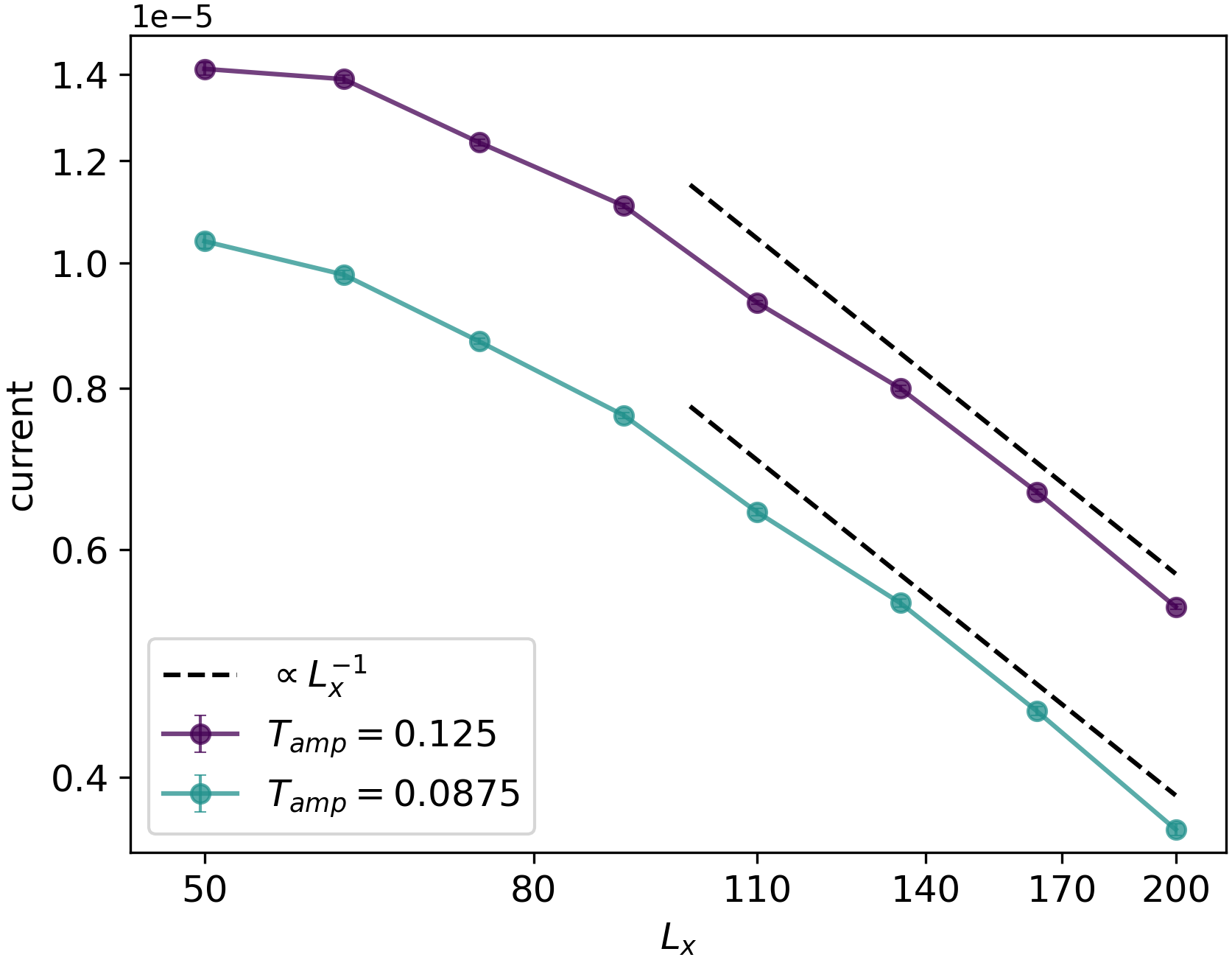}
    \caption{Particle current in the $x$ direction as a function of the system size $L_x$ for fixed aspect ratio $L_y/L_x$ and filling factor $\overline\rho = 1/2$. 
    The temperature varies along $x$ as in Eq.~\eqref{eq: x temperature profile} for $T_{\mathrm{mean}}=0.42$, and $T_{\mathrm{amp}}=0.12$ (purple) or $T_{\mathrm{amp}}=0.087$ (green). Average over 40 samples and over the last $25\%$ of the total time $t=1.33 \times 10^7$. See main text.} 
    \label{fig: scaling currents}
\end{figure}

The results are shown in Fig.~\ref{fig: scaling currents} for values of $T_{\mathrm{mean}}$ and $T_{\mathrm{amp}}$ such that the temperature remains subcritical throughout the system. 
The data indicate that the expected $1/L_x$ decay has not yet been reached at the accessible system sizes. 
Nevertheless, the pronounced downward curvature relative to a pure power-law fit is consistent with this asymptotic behavior. 
While the absolute magnitude of the measured current depends on details of the measurement protocol, the overall scaling behavior is robust. 
For instance, probing the current less deeply inside the stripes by replacing the fraction $1/8$ introduced above with $1/2$ yields power-law fits that differ by only a few percent. 
Additional measurements for temperature profiles that are not everywhere subcritical are reported in Appendix~\ref{sec: scaling particle current additional}.

\subsection{Filling Factor $\overline\rho\ne1/2$}\label{sec: other filling factors}

We now examine how the phenomenology changes for filling factors $\overline{\rho}>1/2$.
We focus on systems with $L_x = L_y/2 = 200$ and the temperature profile of Eq.~\eqref{eq: x temperature profile}.

A stationary density profile for $\overline{\rho}=0.8$ is shown in the right panel of Fig.~\ref{fig:equil and non equil}.
Stripes form in the coldest region of the system.
As mentioned in the summary of the results,
this steady-state is markedly different from the equilibrium state obtained for the rates in Eq.~\eqref{eq: equilibrium rates} under the same temperature profile, shown in the left panel of Fig.~\ref{fig:equil and non equil}.
A related contrasting behavior was reported in a one-dimensional model in Ref.~\cite{Bochers_2014}.
At the dynamical level, the key difference is that in the nonequilibrium case defects propagate diffusively within each phase,  while this motion is suppressed for the equilibrium dynamics, with defects being kept away from the cold temperature part, as is visible in the left panel of Fig.~\ref{fig:equil and non equil}; see Sec.~\ref{sec: macroscopic constraints}.


A further indication of nonequilibrium behavior is the geometry of the interfaces: they deviate strongly from the minimal-curvature shapes expected in equilibrium.
The stripes are elongated rather than circular, with even a dumbbell-like profile. 
This reflects the fact that interface shapes are controlled by nonequilibrium currents that dominate over much weaker mean-curvature ones, see Sec.~\ref{sec: interface curvature}.

In Sec.~\ref{sec: chemical potential profiles}, we derive the macroscopic chemical-potential profile $\mu(\mathbf x)$.
This quantity is simpler to analyze than $\rho(\mathbf x)$ because it is insensitive to interfaces.
For the temperature profile as in Eq.~\eqref{eq: x temperature profile}, $\mu$ is expected to be $y$-independent.
The theory in Sec.~\ref{sec: chemical potential profiles} yields a direct prediction for the vertically averaged density
\begin{equation}\label{eq: vertically averaged density formula}
\widetilde \rho(x) = \frac1{L_y}\sum_y \rho(x,y),
\end{equation}
see Eq.~\eqref{eq: solution mu profile} as well as the discussion at the end of Sec.~\ref{sec: chemical potential profiles}.
For $\overline{\rho}>1/2$, $\widetilde{\rho}(x)$ should equal $1/2$ on a symmetric interval around $x=3L_x/4$, and equal the spontaneous density at the interval’s boundary elsewhere.
The width of this plateau is fixed by the value of $\overline{\rho}$.

\begin{figure}[t]
    \centering
    \includegraphics[width=0.95\linewidth]{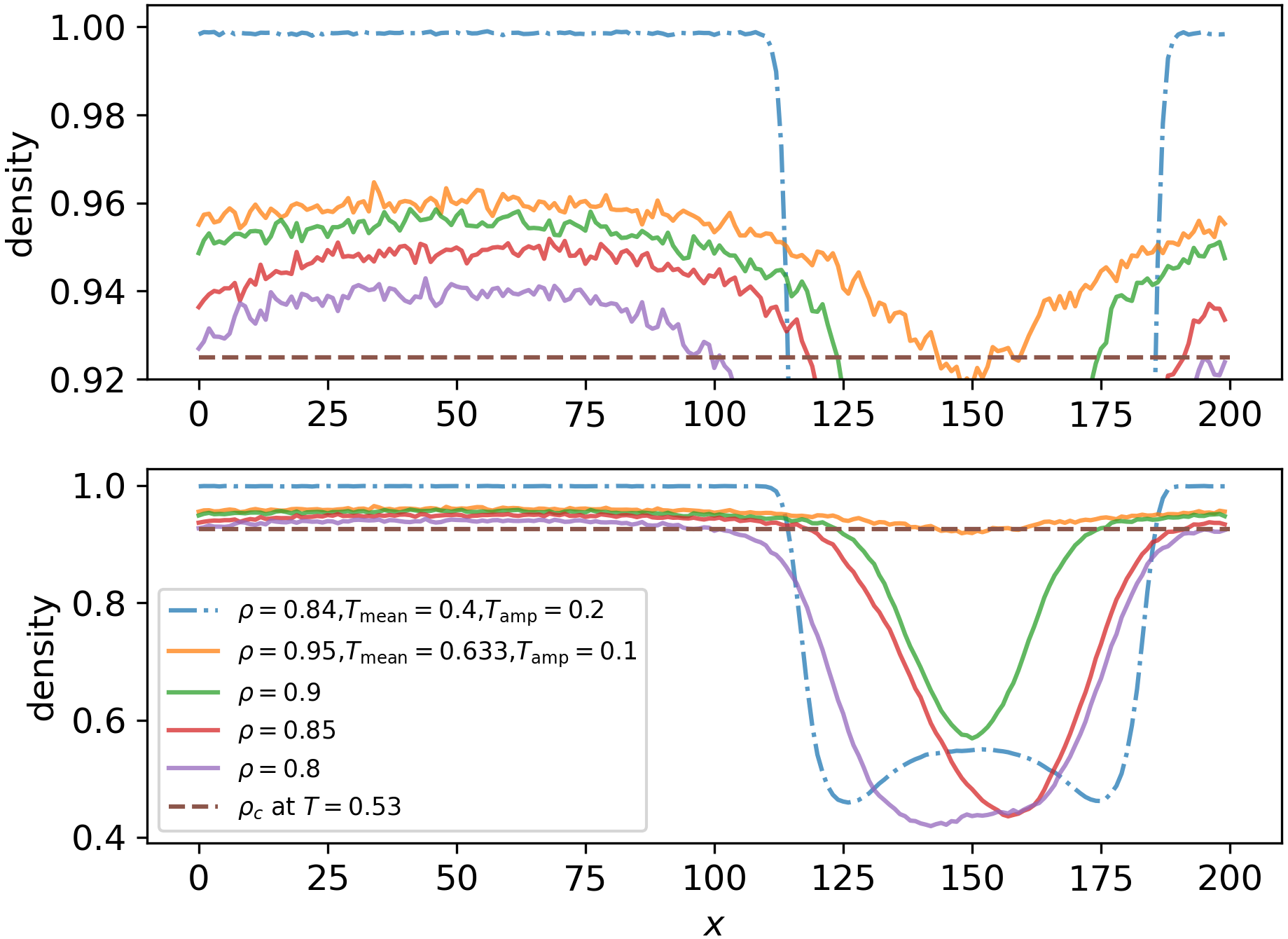}
    \caption{
    Time and $y$-averaged density profile $\tilde{\rho}(x)$ at filling factor $\overline{\rho}=0.8$. 
The temperature varies along $x$ as in Eq.~\eqref{eq: x temperature profile}, with $T_{\mathrm{mean}}=0.4$ and $T_{\mathrm{amp}}=0.2$ (blue dashed line), and $T_{\mathrm{mean}}=0.633$ and $T_{\mathrm{amp}}=0.1$ (solid lines). 
The horizontal brown dashed line indicates the critical density $\rho_c(T=0.53)$. 
The system size is $L_y=2L_x=400$. 
Time is averaged over $75\%$ of the total simulation time $t=5\times 10^7$ for a single realization.
The lower panel shows the full profile, while the upper panel highlights the region where $\tilde{\rho}(x)>\rho_c(T=0.53)$.}
    \label{fig: density at non-half filling}
\end{figure}

Figure~\ref{fig: density at non-half filling} shows $\widetilde{\rho}(x)$ for $\overline{\rho}=0.8$ and various $T_{\mathrm{mean}}$ and $T_{\mathrm{amp}}$.
The dashed blue curve corresponds to $\widetilde\rho(x)$ for the stationary density in the right panel of Fig.~\ref{fig:equil and non equil}.
Here the spontaneous density at the stripe boundary is very close to $1$, and the agreement with the macroscopic prediction outside the stripe region is good.
Inside the stripe region, however, $\widetilde{\rho}(x)$ varies noticeably around $1/2$, consistent with the dumbbell-like stripe shape.
It remains unclear whether this variation persists at larger system sizes.

We also explore temperature profiles yielding smaller spontaneous densities in the cold region, giving rise to the remaining curves in Fig.~\ref{fig: density at non-half filling}.
A notable observation is that $\widetilde{\rho}(x)$ systematically exceeds the spontaneous density $\rho_c$ in the coldest part by a few percent.
As far as we could test, this effect does not diminish with increasing system size, and we currently do not have an explanation for it.

\section{Macroscopic Description}\label{sec: macroscopic description}

We now turn to a macroscopic description of the nonequilibrium system introduced in Sec.~\ref{sec: Kawasaki out of equilibrium}. 
The framework relies on two fundamental physical principles: local equilibrium in the steady-state and diffusive transport of the conserved quantity (Section~\ref{sec: macroscopic constraints}). 
This setup yields three key results. 
First, out-of-equilibrium steady currents, rather than interface curvature, primarily shape the out-of-equilibrium interfaces (Section~\ref{sec: interface curvature}). 
Second, it determines the chemical potential profiles, matching our numerical observations (Section~\ref{sec: chemical potential profiles}). 
Finally, it explains the physical mechanism behind the observed proliferation of stripes in the large-volume limit (Section~\ref{sec: stripe formation}).

Before moving on, let us note that the temperature profiles considered so far may induce symmetries that are not broken microscopically. 
For example, if the temperature depends only on \(x\) as in Eq.~\eqref{eq: x temperature profile}, the stationary density remains homogeneous in \(y\), reflecting the unfixed vertical position of the stripes. 
As we saw though, particle configurations exhibit well-defined stripes whose positions are remarkably stable, with motion timescales that grow rapidly with system size. 
The present macroscopic theory retains only currents scaling as $1/L$ or larger, thus neglecting dynamics occurring on timescales beyond diffusion. 
As a result, it can treat as stationary configurations that are microscopically quasi-stationary, rather than their broad superpositions.
Accordingly, throughout this section, the density $\rho$ refers to that of a locally pure state, not of a mixture.

Throughout this section, the system size is rescaled by \(1/L\) so that \(L_x/L = a\) and \(L_y/L = b\) for fixed constants \(a,b>0\) as $L\to\infty$. 
The corresponding macroscopic coordinates are \(\mathbf{u} = \mathbf{x}/L \in [0,a]\times[0,b]\) (with periodic boundary conditions).

\subsection{Macroscopic Constrains}\label{sec: macroscopic constraints}

We explicitly state the physical principles constraining the macroscopic steady state: 
local equilibrium in the bulk and at the interfaces 
[Eqs.~\eqref{eq: magnetization everywhere} and \eqref{eq: magnetization interfaces}], 
particle number conservation 
[Eqs.~\eqref{eq: model continuity bulk} and \eqref{eq: model continuity interfaces}], 
and macroscopic diffusive transport away from the interfaces
[Eq.~\eqref{eq: diffusive current}].
See also~\cite{Colangeli_2019,Giardina_2020}.

\paragraph{Local equilibrium.}
Local equilibrium holds throughout the system: 
the expectation of any local observable approaches that of an equilibrium system at the local temperature \(T(\mathbf{u})\) as \(L \to \infty\). 
Our numerical results are consistent with local equilibrium, see Sec.~\ref{sec: local equilibrium}. 
This originates from the fact that the rates in Eq.~\eqref{eq: nonequilibrium rates} break detailed balance by an amount that scales as $1/L$ for $L\to\infty$.

Local equilibrium precludes metastable states, implying that for any \(\mathbf{u}\), 
\begin{equation}
\label{eq: magnetization everywhere}
\rho(\mathbf u) \notin (1 - \rho_c(\mathbf u),\rho_c(\mathbf u))
\qquad \text{(everywhere)}
\end{equation}
where
\[
\rho_c  = \rho_c(T(\mathbf{u}))\ge 1/2
\]
is the spontaneous density, as given by Onsager’s formula~\cite{yang1952ising}.
Near interfaces, where the local density averages to \(1/2\), local equilibrium further predicts
\begin{equation}\label{eq: magnetization interfaces}
\rho_+(\mathbf u) = \rho_c(\mathbf u), \quad 
\rho_-(\mathbf u) = 1 - \rho_c(\mathbf u)
\quad \text{(interfaces)}
\end{equation}
with \(\rho_\pm\) denoting the densities on each side of the interface.

\paragraph{Conservative dynamics.}
Since the dynamics conserves the particle number at the microscopic level, Eq.~\eqref{eq: lattice divergence form}, the macroscopic density field likewise satisfies a continuity equation everywhere.
Let \(J(\mathbf{u})\) denote the macroscopic density current at position \(\mathbf{u}\).
In the bulk, away from interfaces separating high and low-density phases,
\begin{equation}\label{eq: model continuity bulk}
\nabla \cdot J(\mathbf{u}) = 0
\qquad \text{(bulk),}
\end{equation}
while at interfaces the condition reduces to the static Stefan relation,
\begin{equation}\label{eq: model continuity interfaces}
\left(J_+(\mathbf{u}) - J_-(\mathbf{u})\right) \cdot \mathbf{n} = 0
\qquad \text{(interface),}
\end{equation}
where \(J_\pm\) are the currents on each side of the interface and \(\mathbf{n}\) is the unit normal vector.

\paragraph{Fick's law.}
The microscopic continuity equation is given in Eq.~\eqref{eq: lattice divergence form}, with the corresponding microscopic in Eq.~\eqref{eq: nonequilibrium current}. 
Naively interpolating this expression at the macroscopic level yields the following expression for the macroscopic current \(J = (J_x, J_y)\):
\begin{equation*}
J_z = -\partial_z \rho + 2\rho(1-\rho)\tanh\!\big[(5/2T)\,\partial_z \rho\big],
\end{equation*}
$z=x,y$, where we have set $4J=1$, as previously.
Crucially, this current vanishes when \(\nabla\rho = 0\), indicating that the large-scale dynamics is diffusive.
However, we do not expect the diffusion coefficient to be read directly from this expression; instead, we allow for a more general constitutive relation of the form
\begin{equation}\label{eq: diffusive current}
J(\mathbf{u}) = -D(T(\mathbf{u}), \rho(\mathbf{u}))\,\nabla \rho(\mathbf{u})
\qquad \text{(bulk),}
\end{equation}
i.e., Fick’s law, where \(D\) denotes the diffusion coefficient.
The coefficient \(D(T,\rho)\) remains finite and bounded away from zero as long as \(\rho(\mathbf{u}) \ge \rho_c(\mathbf{u})\) and as we keep away from the critical point. 
This is consistent with the observation that, at the microscopic level and in the low temperature region, defects diffuse freely within the majority phase in first good approximation. 
See also~\cite{SpohnYau1995}.
Equation~\eqref{eq: diffusive current} is expected to hold away from interfaces.

Note that the current in Eq.~\eqref{eq: equilibrium current} for the equilibrium dynamics with an inhomogeneous Hamiltonian does not yield diffusive dynamics, since $\Delta_{\mathbf x, \mathbf y}^T$ in Eq.~\eqref{eq: energy difference equilibrium} does not vanish at constant density.

\subsection{Role of Interface Curvature}\label{sec: interface curvature}

Interface curvature plays no role in the macroscopic description presented above. We now explain why these curvature effects, which are vital in shaping equilibrium interfaces, become secondary to out-of-equilibrium currents when temperature gradients are imposed.

In equilibrium at uniform subcritical temperature, with a global filling factor between \(1-\rho_c\) and \(\rho_c\), steady profiles consist of a single bubble of the minority phase with minimal curvature, and the density is critical everywhere. 
Equations~\eqref{eq: magnetization everywhere}--\eqref{eq: diffusive current} are then satisfied. 
However, these conditions alone do not select the observed profiles. 
This reflects the absence of any term involving interface curvature in our macroscopic equations.

Such curvature terms are excluded because only currents of order \(1/L\) are retained in the present description. 
In a more refined treatment, the interfacial densities in Eq.~\eqref{eq: magnetization interfaces} acquire corrections proportional to the inverse radius of curvature, i.e., of order \(1/L\). 
These small variations generate currents scaling as \(1/L^2\), giving rise to slow mean-curvature motion~\cite{bray_1994}.

When the temperature profile is not uniform, microscopic steady currents appear that scale as \(1/L\), see Sec.~\ref{sec: scaling current}. 
Because these dominate over mean-curvature contributions, curvature effects play little role in shaping macroscopic nonequilibrium steady profiles, see e.g.\@ the right panels of the right panel of Figs.~\ref{fig:equil and non equil} and \ref{fig: densities and currents}, where interface shapes clearly deviate from mean-curvature behavior.

\subsection{Chemical Potential Profiles}\label{sec: chemical potential profiles}

Under local equilibrium, we can define the chemical potential $\mu(\mathbf u)$ through 
\[
    \rho(\mathbf{u}) = \langle n \rangle_{T(\mathbf{u}),\mu(\mathbf{u})}.
\]
Determining the chemical potential profile provides more fundamental information than the density profile, as it does not require identifying the position and shape of the interfaces within the $\mu = 0$ region.
In fact, the macroscopic theory allows us to determine it completely, and the result can be described as follows. 
In general, an isothermal line at a subcritical temperature can separate the domain into two regions. 
In the region where the temperature is higher than this line, denoted as the set $\Omega_0$ in Eq.~\eqref{eq: Omega not set} below, the density is constant and equal to the spontaneous density on the isothermal line.
In the complement of $\Omega_0$, where the temperature is lower, $\mu = 0$, and this region may be filled with stripes. 
The position of this isothermal line is dictated by the overall density. 
Of course, degenerate cases may occur where no such line exists, meaning $\Omega_0$ is either empty ($\mu=0$ everywhere, as in all our simulations at half filling) or occupies the entire space (which occurs when the density exceeds a specific threshold).

Let us derive these claims.
Combining Eqs.~\eqref{eq: model continuity bulk} and \eqref{eq: diffusive current} gives
\begin{equation}\label{eq: div J = 0 in bulk}
\nabla \cdot J =
\nabla \cdot (D(T(\mathbf{u}),\rho(\mathbf{u}))\,\nabla \rho(\mathbf{u})) = 0,
\end{equation}
valid away from interfaces. 
The only free parameter is the overall filling factor \(\overline{\rho}\), which by symmetry can be taken in \([1/2,1]\).

Using the chain rule,
\(\nabla \rho = (\partial\rho/\partial T)\nabla T + (\partial\rho/\partial \mu)\nabla\mu\),
Eq.~\eqref{eq: div J = 0 in bulk} can be reformulated as an equation for \(\mu\).
When \(\mu \ne 0\), the system is in a single phase and Eq.~\eqref{eq: div J = 0 in bulk} applies.
We impose that \(\mu\) varies continuously across the boundary separating the \(\mu\neq 0\) and \(\mu=0\) regions.

This is sufficient to determine \(\mu(\mathbf{u})\) throughout the system.
Let \(\mathbf{u}_0\) denote a point of maximal temperature and define
\begin{equation}\label{eq: Omega not set}
\Omega_0 = \{ \mathbf{u} : \rho_c(T(\mathbf{u})) < \rho(\mathbf{u}_0) \}.
\end{equation}
We then set
\begin{equation}\label{eq: solution mu profile}
\mu(\mathbf{u}) =
\begin{cases}
\text{s.t. } \langle n \rangle_{T(\mathbf{u}),\mu(\mathbf{u})} = \rho(\mathbf{u}_0), & \mathbf{u}\in\Omega_0,\\
0, & \mathbf{u}\notin\Omega_0.
\end{cases}
\end{equation}
Inside \(\Omega_0\), eq.~\eqref{eq: solution mu profile} is re-expressed more simply as 
\[
\rho(\mathbf{u})=\rho(\mathbf{u}_0).
\]
Hence \(\nabla\rho(\mathbf{u})=0\), satisfying Eq.~\eqref{eq: div J = 0 in bulk}.
On the boundary of \(\Omega_0\), the definition of $\Omega_0$ ensures \(\rho(\mathbf{u}) = \rho_c(T(\mathbf{u}))\), so that \(\mu=0\) there, and $\mu(\mathbf u)$ is continuous everywhere.

Two limiting cases are immediate.
If \(\Omega_0\) spans the whole volume, then \(\rho(\mathbf{u})\) is constant, which occurs when the temperature is everywhere supercritical or when the filling factor satisfies \(\overline{\rho}>\rho_c(T_{\min})\), with \(T_{\min}\) the minimum temperature in the system.
Conversely, if \(\Omega_0=\varnothing\), then \(\mu=0\) everywhere, corresponding to half-filling.

We now consider a temperature profile depending only on \(x\), as in Eq.~\eqref{eq: x temperature profile}, with aspect ratio \(L_y/L_x=2\) such that the macroscopic domain is \([0,1]\times[0,2]\).
For this geometry, \(\Omega_0 = I_0\times[0,2]\), where \(I_0\) is an interval centered around \(x=1/4\) in macroscopic units (periodic boundary conditions).
The width of \(I_0\) is set by the filling factor \(\overline{\rho}\).
The theory predicts that \(\rho(\mathbf{u})\) remains constant within \(\Omega_0\), with a value equal to the spontaneous density at the boundary of \(I_0\) (\(\mu=0\)).
This prediction is confirmed by the observations on Fig.~\ref{fig: density at non-half filling}, up to some deviations discussed in Sec.~\ref{sec: other filling factors}.
Further, it provides a full explanation to the main difference between the left and right panels of Fig.~\ref{fig:equil and non equil}.

\subsection{Stripe Formation}\label{sec: stripe formation}

\begin{figure*}[t]
    \centering
    \includegraphics[width=0.99\linewidth]{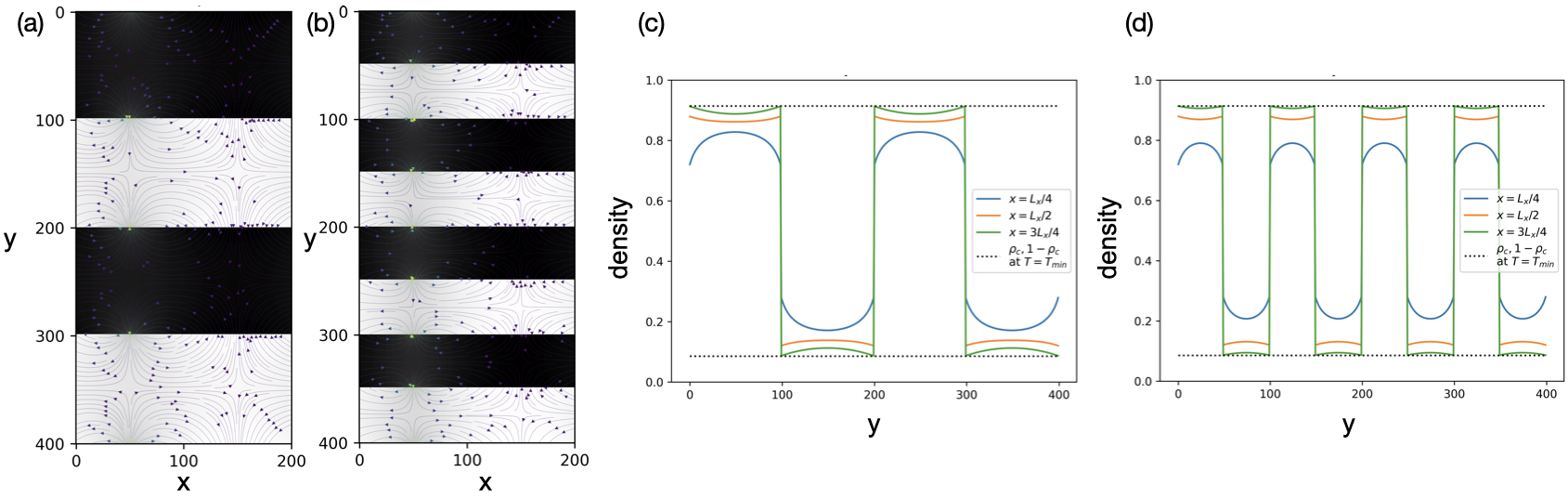}
    \caption{
    Two steady density profiles satisfying Eqs.~\eqref{eq: magnetization interfaces}--\eqref{eq: diffusive current}. 
    Panels (a) and (b) show the spatial density (greyscale) and particle currents (colored arrows), featuring two and four high-density stripes, respectively. 
    Panels (c) and (d) show the density along the $y$ direction at $x=L_x/4$ (blue), $x=L_x/2$ (orange), and $x=3L_x/4$ (green); panel (c) corresponds to the solution in (a), and panel (d) to the one in (b). 
    The horizontal dotted lines mark the spontaneous density in the coldest region ($x=3L_x/4$).}
    \label{fig: macroscopic stripe profiles}
\end{figure*}

We now address the emergence of stripe patterns. 
The central physical insight is that, at the macroscopic level, a point of minimal, subcritical temperature cannot lie within a macroscopic region of high or low density. 
This restriction underlies the formation of stripes observed for the geometry predominantly considered here, the number of which must grow unbounded with the system size. 

Below, we formulate this property as a mathematical fact and discuss its implications for interpreting our numerical results. 
Before doing so, however, let us present an explicit construction to illustrate the mechanism most directly.

\paragraph{Explicit construction of steady profiles.}
\phantom{}
In Fig.~\ref{fig: macroscopic stripe profiles}, we construct profiles that exactly satisfy the macroscopic constraints in Eqs.~\eqref{eq: magnetization interfaces}--\eqref{eq: diffusive current} but (weakly) violate the local equilibrium condition in Eq.~\eqref{eq: magnetization everywhere} near the minimal temperature region; see the green line in panels (c) and (d), which displays a density within the forbidden interval $(1 - \rho_c, \rho_c)$. 
As the number of stripes increases [compare panel (c) with panel (d)], meaning the single-phase domains around the lowest temperature line become less macroscopic, this violation diminishes. 
We identify this as the underlying mechanism driving the proliferation of stripes with increasing system size.

Note also that the stripe configurations in Fig.~\ref{fig: macroscopic stripe profiles} are exactly periodic, ensuring that currents are balanced at interfaces, i.e., that eq.~\eqref{eq: model continuity interfaces} is satisfied. 
This periodicity is consistent with the patterns observed in microscopic simulations, see Fig.~\ref{fig: structure factor} and the discussion in Sec.~\ref{sec: regularity patterns}.

For the simulations in Fig.~\ref{fig: macroscopic stripe profiles}, we set $D=1$ in Eq.~\eqref{eq: diffusive current} for simplicity.
The steady state is obtained by integrating the spatially discretized heat equation on a rectangular lattice $\Lambda$ of $L_x \times L_y = 200 \times 400$ sites. 
The system is evolved long enough to reach stationarity within each domain delimited by a phase interface [see panels (a) and (b) in Fig.~\ref{fig: macroscopic stripe profiles}]. 
For each of them, we impose fixed boundary conditions along the interfaces, where the density equals the positive or negative spontaneous density at the corresponding temperature.

\paragraph{Rigorous foundation.}
Going beyond these specific simulations, we show that a point of minimal, subcritical temperature cannot lie within a macroscopic high- or low-density region. 
This restriction is a direct consequence of the tension between local equilibrium on the one hand [Eqs.~\eqref{eq: magnetization everywhere} and \eqref{eq: magnetization interfaces}] and macroscopic diffusive transport on the other hand [Eq.~\eqref{eq: diffusive current}].

Let us consider temperature profiles as in our simulations, i.e.\@ as in Eq.~\eqref{eq: x temperature profile} or Eq.~\eqref{eq: radial temperature profile}, though the claim is more general.
We also assume that the profile stays subcritical throughout, as crossing the critical temperature may lead the diffusion coefficient $D$ in \eqref{eq: diffusive current} to vanish. 

\begin{claim}
Assume that \(\rho(\mathbf{u})\) is a steady macroscopic density profile satisfying Eqs.~\eqref{eq: magnetization everywhere}--\eqref{eq: diffusive current}. 
Then either \(\rho(\mathbf{u})\) is constant, or no point where the temperature is minimal can lie in the interior of a high or low-density region.
\end{claim}

A formal proof is deferred to Appendix~\ref{sec: proof of claim}; here, we outline the core of the argument. 
The primary analytical tool is the maximum principle (see, e.g., Ref.~\cite{gilbarg_trudinger}) applied to the steady state of our diffusive system [Eqs.~\eqref{eq: model continuity bulk} and \eqref{eq: diffusive current}]. 
This principle states that a steady density profile $\rho(\mathbf u)$ must reach its maximum on the boundary of any domain that does not cross any phase interface (i.e., where the high-density phase transitions to the low-density phase).

We assume that $\rho(\mathbf u)$ is not constant and, to establish a contradiction, suppose there exists a domain containing a point of minimal temperature in its interior, with the phase interface as its boundary.
By Eq.~\eqref{eq: magnetization interfaces}, the density on this boundary must equal the critical density. 
However, the local equilibrium constraint in Eq.~\eqref{eq: magnetization everywhere} requires that the density $\rho$ at the coldest point be greater than or equal to the spontaneous density $\rho_c$ at that specific temperature. 
Because $\rho_c(T)$ increases monotonically as temperature decreases, $\rho$ at the coldest point must be larger than the critical density value fixed on the warmer boundary. 
This requirement implies that the maximum is reached in the interior of the domain, yielding a contradiction.

\paragraph{Implications for microscopic simulations.}

We examine how this constraint is reflected, or circumvented, in the microscopic systems studied numerically. 
For the $x$-dependent temperature profile of Eq.~\eqref{eq: x temperature profile}, the minimum temperature lies along $x=3L_x/4$. 
For filling factors close to $1/2$, interfaces must form, and the observation above implies that, in the macroscopic limit, no point on this line can lie deep inside a single phase. 
Microscopic simulations indeed show that the system fragments into an increasing number of stripes as the system size grows,see Fig.~\ref{fig: number of stripes}. 
Our observation implies that the number of stripes must indeed increase with system size, so that no point becomes truly internal to a phase in the macroscopic limit. 
Similar reasoning applies to the “Mexican-hat’’ profiles in Eqs.~\eqref{eq: radial temperature profile} and~\eqref{eq: first radial profile}.

\section{Discussion and Outlook}

We have studied how the phase-separated low-temperature state of a lattice gas with Ising interactions is modified when driven out of equilibrium by a macroscopic temperature gradient, in a regime where local equilibrium is maintained throughout the system. 
We find that, at the macroscopic scale, the ordered phase reorganizes into a regular array of convection cells whose geometry is remarkably robust in time. 
These nonequilibrium patterns differ markedly from the steady states obtained under equilibrium dynamics at the corresponding local temperatures, highlighting the nontrivial impact of weak nonequilibrium driving.
The observations can be qualitatively explained within a macroscopic, or hydrodynamic, description of the system. 

Our results raise several questions concerning the nature of nonequilibrium steady states and their fluctuations. 
E.g., can one develop a quantitative theory of macroscopic fluctuations that is sufficiently predictive to account for the scaling of the number of current vortices? 
Are the observed patterns and their stability specific to open systems, here realized by coupling to a heat bath throughout the bulk, or do analogous structures arise in closed systems? 
Can the shape and organization of the steady-state profiles be understood in terms of fundamental properties of entropy production and, more fundamentally, could they point to organizing principles that replace free-energy minimization in nonequilibrium settings? 
We leave these questions and many others for future work.

\section*{Acknowledgments}

F.H. warmly thanks T.~Demaerel and W.~De Roeck for numerous discussions and exchanges of ideas on the topic of this paper, B.~Derrida for suggesting the thought experiment leading to Fig.~\ref{fig:equil and non equil}, as well as C.~Von Keyserlingk, A.~Nahum and V.~Oganesyan for discussions.  
K.A. thanks K.~Kawaguchi and N.~Nakagawa for insightful comments.
This work was supported by JSPS KAKENHI Grant Numbers JP26K00052 and JP26K17111 (to K.A.).

\bibliography{bibliography_stripes}

\clearpage
\appendix

\section{Alternative Dynamics}\label{sec: other dynamics}

To stress the robustness of our main findings, we consider three variants of the dynamics studied in the main text:

\begin{enumerate} 
    \item 
    The dynamics is as described in Sec.~\ref{sec: Kawasaki out of equilibrium}, but the temperature variation in Eq.~\eqref{eq: x temperature profile} is replaced by a sinusoidal inverse temperature profile:
    \begin{equation}\label{eq: x temperature inverse} 
    \frac{1}{T(u_x,u_y)} = \beta_{\mathrm{mean}} - \beta_{\mathrm{amp}} \sin (2 \pi u_x) 
    \end{equation} 
    with $\beta_{\mathrm{mean}} = 1.4 \beta_c$ and $\beta_{\mathrm{amp}} = 0.5 \beta_c$ ($\beta_c \approx 1.76$).

    \item 
    The dynamics is as described in Sec.~\ref{sec: Kawasaki out of equilibrium}, but the temperature variation in Eq.~\eqref{eq: x temperature profile} is replaced by a sawtooth profile:
    \begin{multline}\label{eq: saw temperature profile}
        T(u_x,u_y) = \\
        \begin{cases}
            4 T_{\mathrm{amp}} u_x + T_{\mathrm{mean}}, 
            & 0 \le u_x \le 1/4,\\
            - 4 T_{\mathrm{amp}} (u_x - 1/2) + T_{\mathrm{mean}}, 
            & 1/4 \le u_x \le 3/4, \\
             4 T_{\mathrm{amp}} (u_x-1) + T_{\mathrm{mean}}, 
            & 3/4 \le u_x \le 1, 
        \end{cases}
    \end{multline}
    with $T_{\mathrm{mean}}=0.4$ and $T_{\mathrm{amp}} = 0.2$. 

    \item 
    The temperature profile remains as defined in Eq.~\eqref{eq: x temperature profile} (with $T_{\mathrm{mean}}=0.4$ and $T_{\mathrm{amp}} = 0.2$), but the rate $\gamma$ in Eq.~\eqref{eq: nonequilibrium rates} is made temperature-dependent:
    \begin{equation}\label{eq: gamma dependent} 
    \gamma_{\mathbf x, \mathbf y} = \left(\frac{T_{\mathbf x, \mathbf y}}{T_{\mathrm{max}}}\right)\gamma 
    \end{equation} 
    where $T_{\mathrm{max}} = T_{\mathrm{mean}} + T_{\mathrm{amp}}$. 
    This ensures $0 < \gamma_{\mathbf x, \mathbf y} \le \gamma$ and models the slowing of dynamics in colder regions.
\end{enumerate} 

\begin{figure*}[t]
    \centering
    \includegraphics[width=0.99\linewidth]{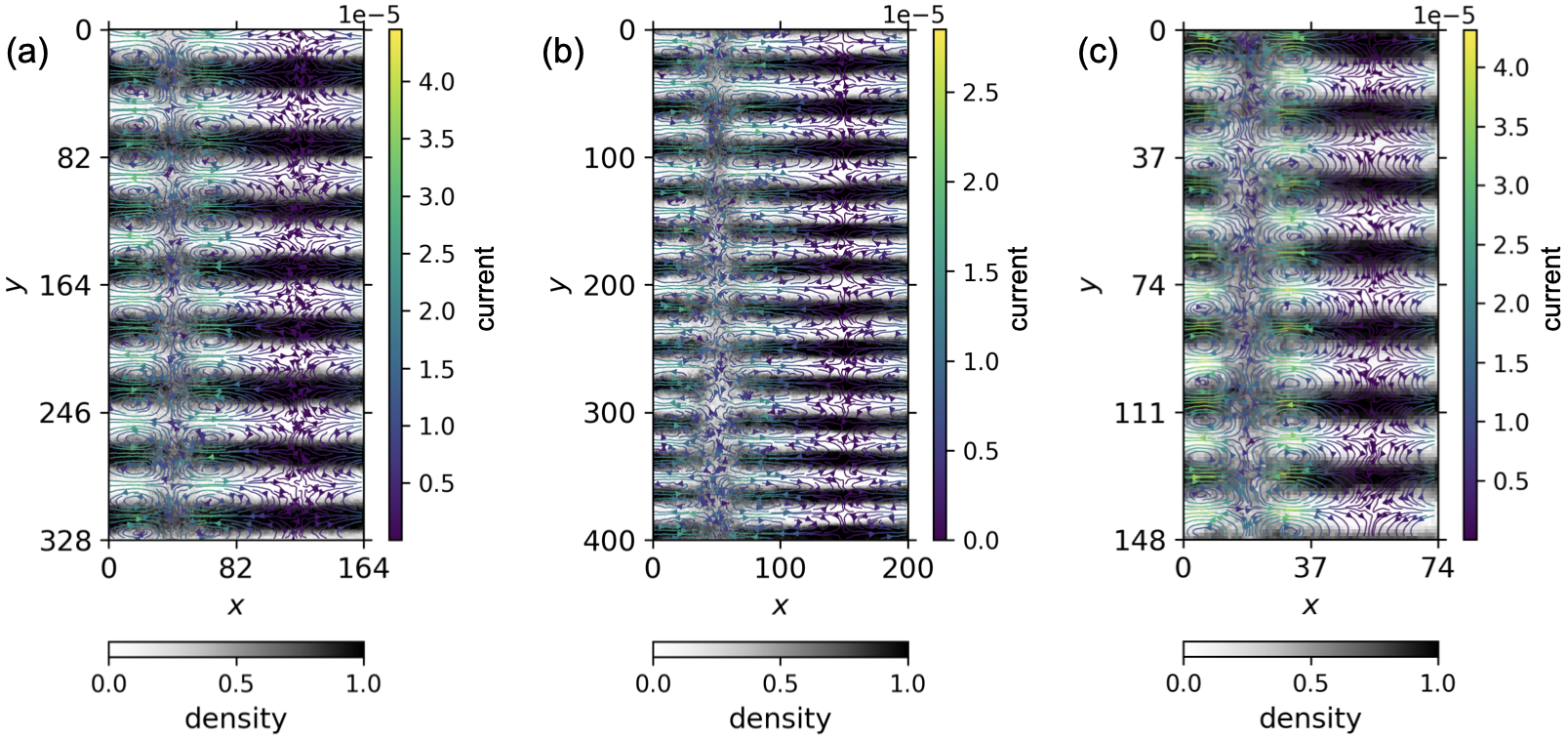}
    \caption{Time-averaged densities (grayscale) and particle currents (colored lines) for a single realization and half filling ($\overline\rho=1/2$). 
    Averaging is performed over the final $25\%$ of the total simulation time ($t = 5 \times 10^{7}$). 
    Panel (a): $x$-dependent temperature profile as in Eq.~\eqref{eq: x temperature inverse} with $L_y = 2L_x = 328$. 
    Panel (b): $x$-dependent temperature profile as in Eq.~\eqref{eq: saw temperature profile} with $L_y = 2L_x = 400$.
    Panel (c): $x$-dependent temperature profile as in Eq.~\eqref{eq: x temperature profile} and temperature dependent $\gamma$ as in~\eqref{eq: gamma dependent}, with $L_y = 2L_x = 148$.
    }
    \label{fig: densities and currents additional}
\end{figure*}

Figure~\ref{fig: densities and currents additional} shows the average density and currents for these cases. 
Averages are taken over the final $25\%$ of a simulation of total duration $t=5\times 10^7$, starting from an infinite-temperature state at half filling ($\overline\rho = 1/2$). 
The aspect ratio is $L_y = 2 L_x$. 
The resulting steady states are qualitatively analogous to that shown on the left panel of Fig.~\ref{fig: densities and currents}, confirming that the salient features of the non-equilibrium steady states are not sensitive to details of the model.

\section{Assessing Stationarity}\label{sec: evidence for stationarity}

We perform two tests to check that the dynamics have reached configurations representative of the stationary state. 
As noted in the main text, once convection cells form, their number may fluctuate over time, but the time scales for these processes are too slow to be seen in our simulations.

\begin{figure*}[h]
    \centering
    \includegraphics[width=0.99\linewidth]{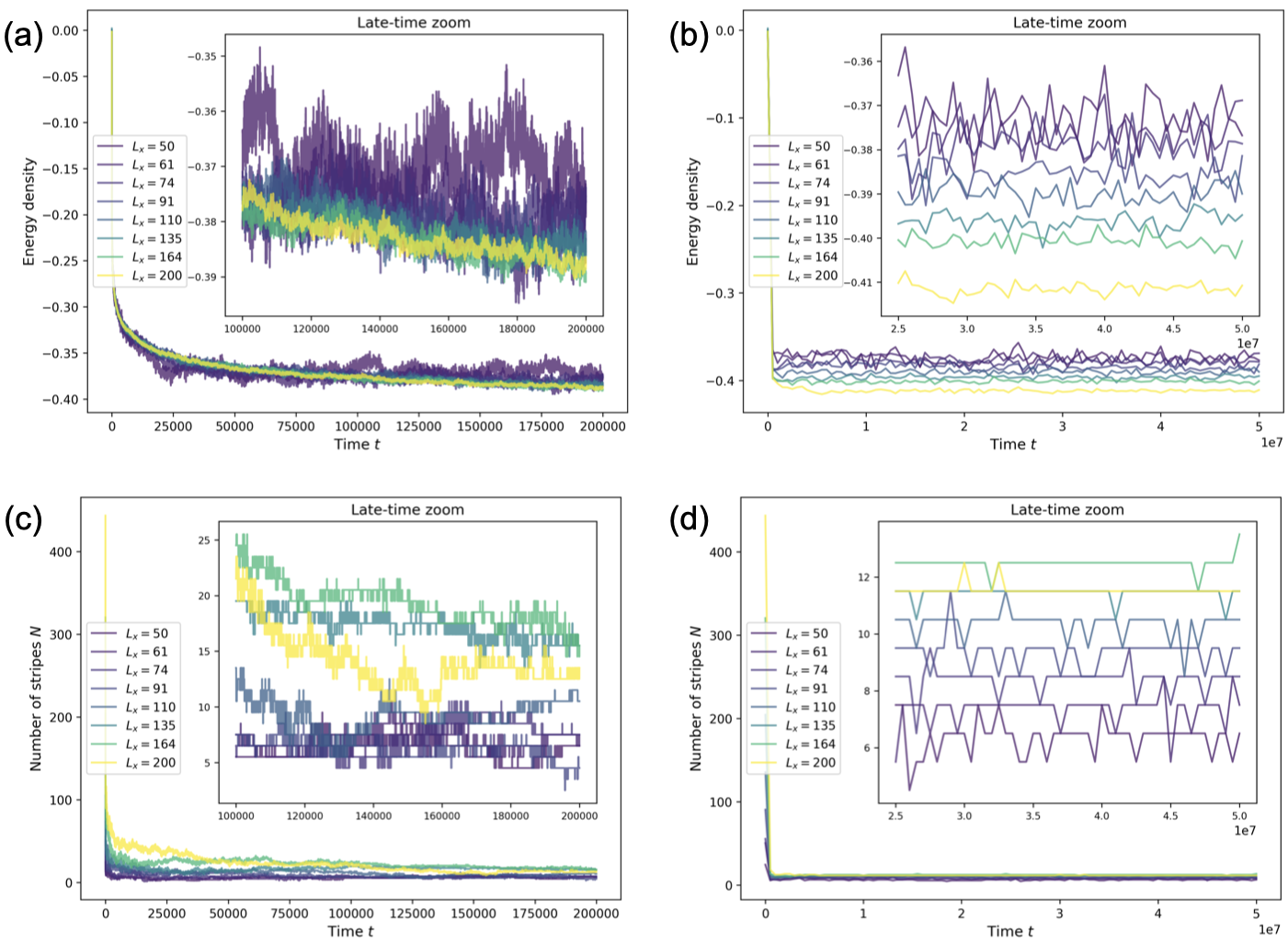}
    \caption{Time dependence of the energy density (top panels) and the number of stripes $\mathcal{N}$ (bottom panels). 
    The data shown are for a single realization starting from an infinite-temperature configuration at half filling ($\bar{\rho} = 1/2$), with the temperature profile given in Eq.~\eqref{eq: x temperature profile} and an aspect ratio $L_y/L_x = 2$. 
    Different system sizes $L_x$ are indicated in the inset. 
    Left panels (a) and (c) show a short time window $[0, 2\times 10^5]$ with $2\times 10^4$ measurements. 
    Right panels (b) and (d) show the long time window $[0, 5\times 10^7]$ with 100 measurements, which corresponds to the window used for the majority of results in the main text.}
    \label{fig: time dependence and stationarity}
\end{figure*}

The first test is to measure the energy density, $E(\eta(t))/V$, with $V=L_xL_y$ and $E(\eta)$ the energy given in Eq.~\eqref{eq: ising energy}. 
This quantity is expected to become approximately constant once the stationary regime is reached. 
The results are shown in the top panels of Fig.~\ref{fig: time dependence and stationarity} for a system with $T_{\mathrm{mean}}=0.4$, $T_{\mathrm{amp}}=0.2$, and aspect ratio $L_y/L_x = 2$. 
Starting from an initial configuration at infinite temperature for every system size considered, the energy density is plotted as a function of time for a short window $[0,2\times 10^5]$ in panel (a), and for a long window $[0,5 \times 10^7]$ in panel (b). 
The latter is the window used for most results in the main text. 
The results in Fig.~\ref{fig: time dependence and stationarity} show that the energy density stabilizes on time scales that are only a few percent of the total simulation time. 

The second test measures how the number of convection cells (or high-density stripes $\mathcal{N}$ as considered in Sec.~\ref{sec: scalcing number of stripes}) varies with time. 
See Appendix~\ref{sec: measuring number of stripes} for details on how this is measured. 
The results are shown in the bottom panels of Fig.~\ref{fig: time dependence and stationarity} for the same system, initial conditions, and time windows as the energy. 
These results also show a stabilization of the number of cells on time scales similar to those of the energy.

\section{Measuring the number of Convection Cells}\label{sec: measuring number of stripes}

The number of convection cells, or number of stripes $\mathcal N$, reported in Figs.~\ref{fig: number of stripes}, \ref{fig: time dependence and stationarity} (c)-(d) and \ref{fig: number of stripes, beyond critical}, is obtained as follows. 
For each particle configuration, we restrict attention to the region $L_x/2 \le x < L_x$, where the temperature is subcritical and stripes form. 
We identify connected components of particles, accounting for periodic boundary conditions in the $y$-direction. 
Connected components containing fewer than $10$ particles are discarded. 
The remaining number of connected components is identified with the number of stripes.

For Figs.~\ref{fig: number of stripes} and \ref{fig: number of stripes, beyond critical}, $100$ configurations are recorded at uniformly spaced times over the full simulation for each realization. 
Only configurations within the final $25\%$ of the simulation time are retained, yielding $25$ configurations per realization. 
The reported values of $\mathcal N$ are averaged over these configurations and over independent realizations.

\section{Number of Convection Cells: Additional Data}\label{sec: scaling number stripes super ciritical}

\begin{figure}[h]
    \centering
    \includegraphics[width=0.99\linewidth]{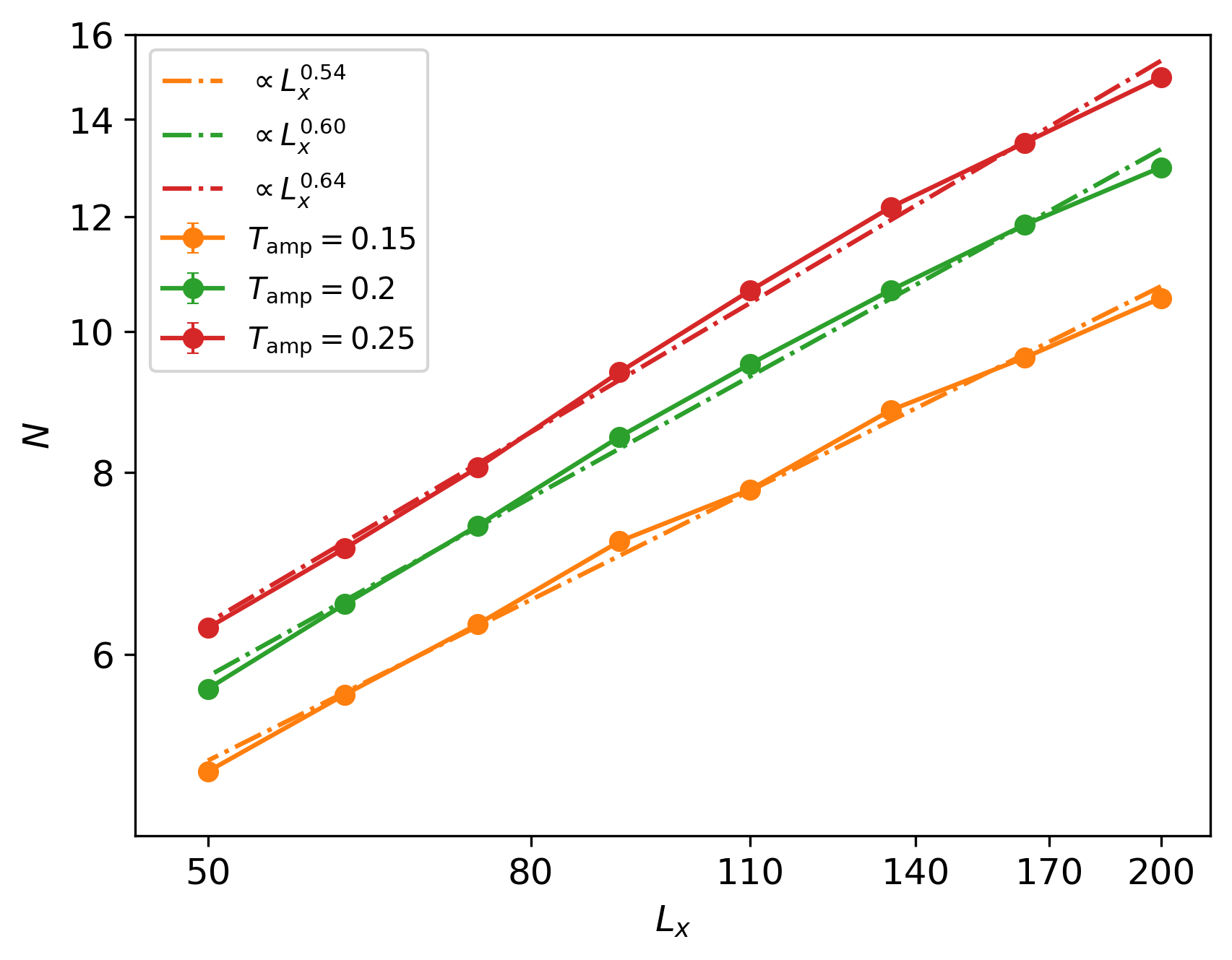}
    \caption{
    Average number of high-density stripes $\mathcal N$ as a function of the system size $L_x$ for fixed aspect ratio $L_y/L_x$ and filling factor $\overline\rho = 1/2$. 
    The temperature varies along $x$ as in Eq.~\eqref{eq: x temperature profile}, with $T_{\mathrm{mean}}=0.4$ and several values of the amplitude $T_{\mathrm{amp}}$ (see legend). 
    Average obtained from $120$ independent realizations and averaged over the final $25\%$ of the simulation time. 
    The total simulation times are $t=5\times10^{6}$ for $T_{\mathrm{amp}}=0.15$, $t=1.33\times10^{7}$ for $T_{\mathrm{amp}}=0.2$, and $t=2.66\times10^{7}$ for $T_{\mathrm{amp}}=0.25$.}
    \label{fig: number of stripes, beyond critical}
\end{figure}

We report additional data for the number of convection cells, or number of stripes $\mathcal N$, obtained with temperature profiles of the form Eq.~\eqref{eq: x temperature profile} that intersect the critical temperature, such that only part of the system is subcritical.
We fix $T_{\mathrm{mean}}=0.2$. 
As for Fig.~\ref{fig: number of stripes} in the main text, we simulate systems with aspect ratio $L_y/L_x = 6$, for $50 \le L_x \le 200$, to mitigate the threshold effect arising form the number of stripes being an integer, and the values for the average number of stripes reported in Fig.~\ref{fig: number of stripes, beyond critical} are obtained by dividing the measured stripe counts by~3 for consistency.
The curve for $T_{\mathrm{amp}}=0.15$ closely matches that obtained in the main text for $T_{\mathrm{mean}}=0.42$ and $T_{\mathrm{amp}}=0.12$, indicating good consistency between the two datasets. 
As expected, the number of stripes increases further with increasing temperature gradient. 
We observe a downward bending of the curves $T_{\mathrm{amp}}=0.2$ and $T_{\mathrm{amp}}=0.25$, suggesting that the asymptotic scaling regime may not yet be reached.

\section{Scaling of the Particle Current: Additional Data}
\label{sec: scaling particle current additional}

\begin{figure}[h]
    \centering
    \includegraphics[width=0.99\linewidth]{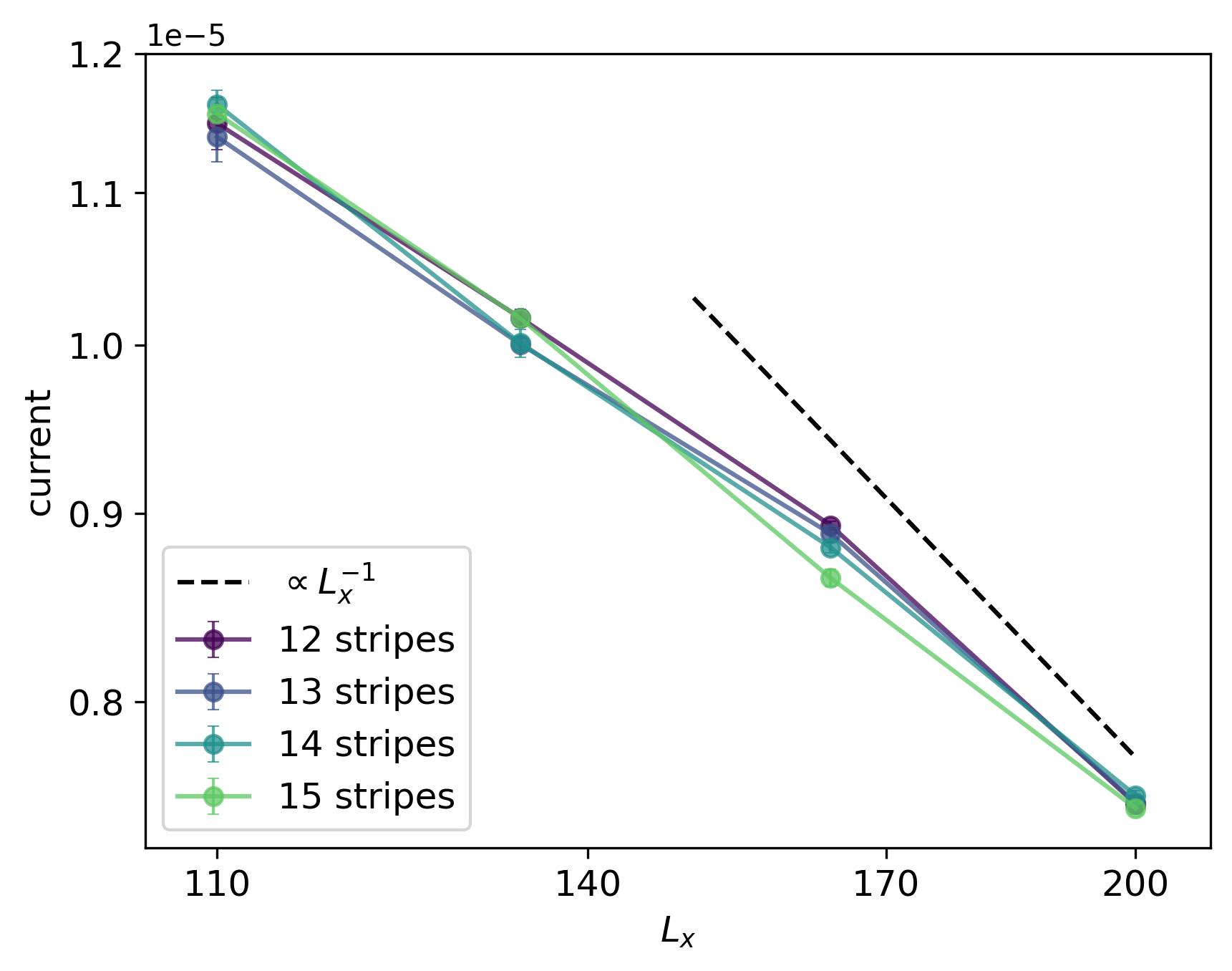}
    \caption{Particle current in the $x$-direction as a function of system size $L_x$ for fixed aspect ratio $L_y/L_x=2$ and filling factor $\overline{\rho}=1/2$. 
    The temperature varies along $x$ as in Eq.~\eqref{eq: x temperature profile}, with $T_{\mathrm{mean}}=0.4$ and $T_{\mathrm{amp}}=0.2$. 
    Simulations are initialized with a fixed number of equally spaced stripes (from $12$ to $15$). 
    Data are averaged over $5$ realizations for each initial condition, and over the final $25\%$ of a simulation of duration $t=2\times10^{8}$.}
    \label{fig: currents through critical}
\end{figure}

We present additional measurements of the particle current in the $x$ direction for the temperature profile of Eq.~\eqref{eq: x temperature profile}, focusing on cases where the temperature is not subcritical throughout the system. 
We examine its scaling with system size for $T_{\mathrm{mean}}=0.4$ and $T_{\mathrm{amp}}=0.2$, where the temperature profile crosses the critical point. 
Results are shown in Fig.~\ref{fig: currents through critical}. 
Simulations are initialized with a prescribed number of equally spaced stripes (ranging from 12 to 15), corresponding to typical steady-state configurations. 
The current is measured using the protocol described in Sec.~\ref{sec: scaling current}. Compared with the data in the main text, obtained for a system where the temperature remains subcritical, the asymptotic $1/L_x$ scaling is not yet observed at the largest sizes simulated. 
Because the profile crosses the critical point, it is unclear if this scaling will eventually be reached, and we draw no firm conclusions from these data.

\section{Proof of the Claim in Section~\ref{sec: stripe formation}}\label{sec: proof of claim}

We provide a more formal proof of the Claim in Section~\ref{sec: stripe formation}.
In the absence of interfaces, Eq.~\eqref{eq: div J = 0 in bulk} holds everywhere, implying that \(\rho(\mathbf{u})\) is constant. 
For a profile with interfaces, let \(\Omega_0 \subset [0,a]\times[0,b]\) be a connected domain occupied by one phase. 
Suppose, for contradiction, that an interior point \(\mathbf{u}_0 \in \Omega_0\) corresponds to a temperature minimum. 
Local equilibrium conditions \eqref{eq: magnetization everywhere} and \eqref{eq: model continuity interfaces}, together with the monotonic increase of \(\rho_c(T)\) with \(T\), give
\[
\rho(\mathbf{u}_0)\ge \rho_c(\mathbf{u}_0) \ge \rho_c(\mathbf{u}) = \rho(\mathbf{u})
\qquad 
\forall \mathbf{u}\in\partial \Omega_0,
\]
where \(\partial\Omega_0\) denotes the boundary of \(\Omega_0\).
Hence \(\rho(\mathbf{u})\) attains its maximum inside \(\Omega_0\). 
Since Eq.~\eqref{eq: div J = 0 in bulk} holds within \(\Omega_0\), the strong maximum principle~\cite{gilbarg_trudinger} implies that \(\rho(\mathbf{u})\) must be constant there. 
For temperature profiles as in our simulations (but it is very generic), the boundary \(\partial\Omega_0\) will include points where \(T(\mathbf{u})>T(\mathbf{u}_0)\), implying \(\rho_c(\mathbf{u}_0)>\rho_c(\mathbf{u})\). 
This implies also $\rho(\mathbf{u}_0)>\rho(\mathbf{u})$, in contradiction with $\rho$ being constant.

\end{document}